\documentclass[10pt,aps,prd,twocolumn,superscriptaddress,amssymb,amsfonts,amsmath,nofootinbib,preprintnumbers,floatfix,secnumarabic]{revtex4-2}
\usepackage{graphicx,slashed,xcolor,multirow,bbold,mathtools,sidecap,tikz,bm,ulem,enumitem,booktabs,array,amsmath,amssymb,slashed}
\usepackage[colorlinks=true,linktoc=page,linkcolor=purple,citecolor=teal,urlcolor=magenta]{hyperref}
\usepackage[compat=1.1.0]{tikz-feynman}
\tikzfeynmanset{warn luatex=false}
\usepackage{siunitx,adjustbox,comment}
\definecolor{lime}{HTML}{A6CE39}
\DeclareRobustCommand{\orcidicon}{\hspace{-1.5mm}
\begin{tikzpicture}
\draw[lime,fill=lime] (0,0.0) circle [radius=0.13] node[white] {{\fontfamily{qag}\selectfont \tiny ID}}; \draw[white,fill=white] (-0.0525,0.095) circle [radius=0.007]; 
\end{tikzpicture} \hspace{-3.0mm} }
\foreach \x in {A, ..., Z} {\expandafter\xdef\csname orcid\x\endcsname{\noexpand\href{https://orcid.org/\csname orcidauthor\x\endcsname} {\noexpand\orcidicon}}}

\let\emph\textit

\DeclareUnicodeCharacter{202F}{\,}

\begin{document}

\preprint{TTP26-024, P3H-26-056}

\title{Searching for charged Higgs bosons in top decays via the \texorpdfstring{$t^*b$}{t*b} channel}

\author{Saiyad Ashanujjaman\orcidA{}}
\email{saiyad.ashanujjaman@kit.edu}
\affiliation{Institut f\"ur Theoretische Teilchenphysik, Karlsruhe Institute of Technology, Engesserstra\ss e 7, D-76128 Karlsruhe, Germany}
\affiliation{Institut f\"ur Astroteilchenphysik, Karlsruhe Institute of Technology, Hermann-von-Helmholtz-Platz 1, D-76344 Eggenstein-Leopoldshafen, Germany}

\author{Guglielmo Coloretti\orcidB{}}
\email{guglielmo.coloretti@bo.infn.it}
\affiliation{INFN, Sezione di Bologna, Via Irnerio 46, 40126 Bologna, Italy}

\author{Andreas~Crivellin\orcidC{}}
\email{andreas.crivellin@cern.ch}
\affiliation{Universitat Autònoma de Barcelona, 08193 Bellaterra, Barcelona}
\affiliation{ICREA, Instituci\'o Catalana de Recerca i Estudis Avan\c{c}ats, Passeig de Llu\'{\i}s Companys 23, 08010 Barcelona, Spain}

\author{Benjamin Fuks\orcidD{}}
\email{fuks@lpthe.jussieu.fr}
\affiliation{Laboratoire de Physique Théorique et Hautes Énergies (LPTHE), UMR 7589, Sorbonne Université et CNRS, 4 place Jussieu, 75252 Paris Cedex 05, France}

\author{Yeonsu Heo\orcidE{}}
\email{yeoheo@student.ethz.ch}
\affiliation{Institute for Theoretical Physics, ETH Z\"urich, Wolfgang-Pauli-Strasse 27, 8093 Z\"urich, Switzerland}

\begin{abstract}
Rare top-quark decays offer a sensitive probe of charged Higgs bosons with masses below the top mass, owing to the large $t\bar t$ production rate at the LHC and the distinctive final states involving leptons and $b$-jets. While existing searches target the $H^\pm\to \tau\nu$, $cs$, and $cb$ modes, the decay $H^\pm\to tb$ has been studied only for heavier charged Higgs bosons with an on-shell top quark in the final state. The low-mass off-shell channel $H^\pm\to t^*b$ therefore remains essentially unconstrained, even though it can become the dominant decay mode below the top-quark threshold owing to the large top Yukawa coupling. We study charged-Higgs production from the rare top decay $t\to H^\pm b$, followed by the decay $H^\pm\to t^*b$. Top-antitop production and decay hence give rise to a $t\bar t b\bar b$-like final state, which we constrain by reinterpreting recent ATLAS fiducial measurements in dileptonic events. We obtain model-independent limits of 1.9\% -- 2.9\% on the product of branching ratios ${\rm Br}(t\to H^\pm b)\times{\rm Br}(H^\pm\to t^*b)$ for charged-Higgs masses between 110 and 165\,GeV, and interpret these bounds in several two-Higgs-doublet scenarios. While dedicated searches for conventional charged-Higgs decays dominate for canonical $Z_2$-symmetric models, the $t^*b$ reinterpretation becomes competitive near the top-quark threshold in the up-type aligned limit and provides relevant direct constraint in a top-philic scenario. In the top-specific Two-Higgs-Doublet Model, it additionally excludes a low-$\tan\beta$ region not covered by conventional searches. These results thus establish $t\bar t$ production with additional $b$-jets as a complementary probe of light charged Higgs bosons and nonstandard top-quark decays, and motivate dedicated analyses by the LHC collaborations. 
\end{abstract}

\maketitle

\section{Introduction}

The top quark plays a central role in tests of the Standard Model (SM) and searches for physics beyond it. As the heaviest known elementary particle, it couples strongly to the electroweak symmetry-breaking sector, while the vast number of top quarks produced at the LHC enables precision studies of rare decay modes~\cite{ATLAS:2024kxj, CMS:2024irj}. Among the most compelling possibilities is the decay $t\to H^\pm b$~\cite{Rizzo:1989ci, CoarasaPerez:1998sqz, Bejar:2000ub}, whose discovery would provide direct evidence for a charged Higgs boson and thus a non-minimal scalar sector.

Searches for light charged Higgs bosons with masses smaller than the mass of the top quark (\mbox{$m_{H^\pm}<m_t$}) have a long history. Early direct searches were first performed at LEP through pair production, \mbox{$e^+e^-\to H^+H^-$}. They resulted in lower bounds on the charged Higgs mass of about 80\,GeV, the exact value depending on the assumed branching fractions into the $cs$ and $\tau\nu$ final states~\cite{ALEPH:2013htx}. At the Tevatron, the CDF and D\O\ collaborations searched for charged Higgs bosons produced in top-quark decays, placing limits on the branching ratio ${\rm Br}(t\to H^\pm b)$ in several benchmark scenarios, using again the $H^\pm\to cs$ and $H^\pm\to \tau\nu$ channels~\cite{CDF:2009efz, D0:2009oou}. More recently, the ATLAS and CMS collaborations have significantly extended the sensitivity using the large top-quark samples collected at the LHC, deriving stringent constraints on the rare $t\to H^\pm b$ decay in the $\tau\nu$~\cite{CMS:2019bfg,ATLAS:2024hya}, $cs$~\cite{CMS:2020osd, ATLAS:2024oqu} and $cb$~\cite{ATLAS:2023bzb} modes. Furthermore, the $WZ$ mode was studied in Ref.~\cite{Ashanujjaman:2025una}. Dedicated searches for charged Higgs bosons decaying into an on-shell top and a bottom quark have also been performed. These analyses target heavy charged Higgs bosons with masses larger than 200\,GeV, produced in association with a top and/or a bottom quark. They hence focus specifically on the on-shell decay $H^\pm\to tb$~\cite{CMS:2019bfg, ATLAS:2021upq, CMS:2025plw}, and are therefore complementary to the light charged-Higgs regime.

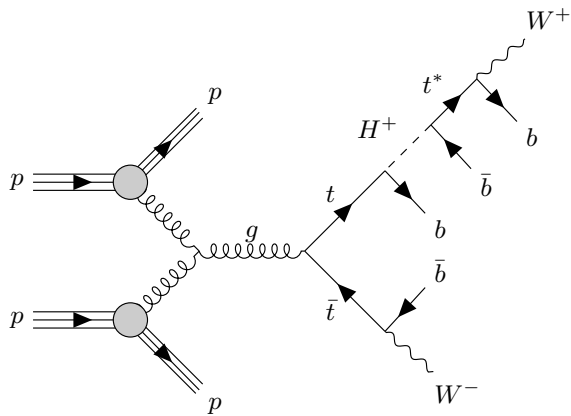
\begin{figure}
  \centering
  \resizebox{0.9\columnwidth}{!}{\begin{tikzpicture}[baseline=(current bounding box.center)]
\begin{feynman}
\vertex (a);

\vertex [above left=1.2cm of a] (c);
\vertex [above=0.1cm of c] (cu); 
\vertex [below=0.1cm of c] (cd);

\vertex [below left=1.2cm of a] (d); 
\vertex [above=0.1cm of d] (du); 
\vertex [below=0.1cm of d] (dd);

\vertex [left=1.2cm of c] (p1) {$p$}; 
\vertex [left=1.2cm of cu] (p1u); 
\vertex [left=1.2cm of cd] (p1d);

\vertex [left=1.2cm of d] (p2) {$p$}; 
\vertex [left=1.2cm of du] (p2u); 
\vertex [left=1.2cm of dd] (p2d);

\vertex [above right=1.2cm of c] (pp1) {$p$}; 
\vertex [above right=1.15cm of cu] (pp1u); 
\vertex [above right=1.26cm of cd] (pp1d); 

\vertex [below right=1.2cm of d] (pp2) {$p$}; 
\vertex [below right=1.26cm of du] (pp2u); 
\vertex [below right=1.14cm of dd] (pp2d); 

\vertex [right=1.3cm of a] (b) ;

\vertex [above right=1.4cm of b] (e);
\vertex [below right=1.4cm of b] (f);

\vertex [above right=0.8cm of e] (j); 
\vertex [above right=0.8cm of j] (m);
\vertex [below right=0.7cm of j] (n) {$\bar b$};
\vertex [below right=0.7cm of e] (i) {$b$};
\vertex [above right=0.7cm of m] (o) {$W^+$};
\vertex [below right=0.7cm of m] (p) {$b$};

\vertex [above right=0.7cm of f] (q) {$\bar b$};
\vertex [below right=0.7cm of f] (r) {$W^-$};

\diagram*{
(p1) -- [fermion] (c) -- [gluon] (a), (p1u) -- (cu), (p1d) --  (cd),
(p2) -- [fermion] (d) -- [gluon] (a),  (p2u) -- (du), (p2d) -- (dd),

(c) -- [fermion] (pp1), (cu) -- (pp1u), (cd) -- (pp1d),
(d) -- [fermion] (pp2), (du) -- (pp2u), (dd) -- (pp2d),

(a) -- [gluon, edge label=$g$] (b),

(f) -- [fermion, edge label=$\bar t$] 
(b) -- [fermion, edge label=$t$] (e),

(q) -- [fermion] (f) -- [boson] (r),

(j) -- [scalar, edge label'=$H^+$] (e) -- [fermion] (i),

(n) -- [fermion] (j) -- [fermion, edge label=$t^*$] (m),
(o) -- [boson] (m) -- [fermion] (p)
};

\end{feynman}

\node[draw, circle, minimum size=12pt, inner sep=0pt, fill=gray!40] at (c) {};
\node[draw, circle, minimum size=12pt, inner sep=0pt, fill=gray!40] at (d) {};

\end{tikzpicture}}
  \caption{Representative Feynman diagram illustrating $t\bar t$ production at hadron colliders, followed by the decay $t \to H^\pm b$ and $H^\pm \to t^*\bar b \to W^\pm b\bar b$, thus leading to a $t\bar{t}b\bar b$-like final state.}
  \label{fig:Feynman}
\end{figure}

However, the above searches do not exhaust the phenomenology of light charged Higgs bosons. In particular, for scenarios in which $m_{H^\pm}<m_t$, a charged Higgs state can decay into a bottom quark and an off-shell top quark, $H^\pm\to t^*b$. Such a mode is generic in two-Higgs-doublet models (2HDM)~\cite{Gunion:1989we, Branco:2011iw, LHCHiggsCrossSectionWorkingGroup:2013rie, Akeroyd:2016ymd, Arbey:2017gmh, Cheung:2022ndq} and triplet extensions of the Higgs sector~\cite{Ashanujjaman:2024lnr}. Below the $tb$ threshold, the decay $H^\pm \to t^*b$ can still lead to characteristic final states containing leptons and multiple $b$-jets. If one top quark in a $t\bar t$ event decays as $t\to H^\pm b$, followed by the subsequent decay $H^\pm\to t^*b\to W^\pm b\bar b$, the event final state is similar to the one originating from top-antitop production in association with two additional $b$-quarks, as illustrated by the Feynman diagram in Figure~\ref{fig:Feynman}. The resulting topology is therefore $t\bar tb\bar b$-like and could manifest as an excess in heavy-flavor-rich top final states rather than as a resonance decaying through a simple two-body decay channel. Associated measurements are already available from the LHC and are commonly interpreted as tests of the SM description of $t\bar tb\bar b$ production. The same fiducial regions are therefore sensitive to rare top decays involving the $H^\pm\to t^*b$ channel, making them a valuable probe of charged Higgs bosons beyond the benchmark scenarios targeted by existing searches.

In this work, we reinterpret recent ATLAS fiducial measurements of dileptonic $t\bar t$ production in association with additional $b$-jets~\cite{ATLAS:2024aht}, and we use them to constrain the product of branching ratios ${\rm Br}(t\to H^\pm b)\times{\rm Br}(H^\pm\to t^*b)$. We derive model-independent limits on this quantity and subsequently interpret them in the up-type aligned, top-philic, top-specific, and canonical $Z_2$-symmetric two-Higgs-doublet scenarios. This comparison highlights the complementarity of the $t^*b$ channel with existing searches for the conventional $cs$, $cb$, and $\tau\nu$ charged-Higgs decay modes. The mass range considered also overlaps with the region where local excesses near 152\,GeV have been reported in associated diphoton production and multi-lepton final states~\cite{Crivellin:2021ubm, Bhattacharya:2023lmu, Ashanujjaman:2024pky, Bhattacharya:2025rfr}. While these hints remain inconclusive, they provide additional motivation to explore complementary probes of low-mass charged Higgs bosons in models with extended scalar sectors~\cite{Banik:2023vxa, Coloretti:2023yyq, Crivellin:2024uhc, Ashanujjaman:2024pky, Fuks:2024qdt}.

The rest of this study is organized as follows. In Sec.~\ref{sec:analysis}, we introduce our simulation tool chain and how we reinterpret the relevant LHC measurements. We analyze our findings in Sec.~\ref{sec:limits}, providing limits for several 2HDM setups, and finally summarize our work in Sec.~\ref{sec:prospects}.

\section{Signal modeling and reinterpretation of \texorpdfstring{$t\bar t b\bar b$}{ttbb} measurements}\label{sec:analysis}

We consider scenarios with charged Higgs bosons lighter than the top quark, their masses satisfying $m_{H^\pm}<m_t$. Such states can be produced in top-antitop events when one of the top quarks undergoes the rare decay $t\to H^\pm b$, followed by the charged Higgs decay $H^\pm\to t^*b$, where the top quark is necessarily off shell. The other top quark of the $t\bar t$ pair decays as in the SM via the decay $t\to Wb$, as shown in Figure~\ref{fig:Feynman}. The resulting signal is therefore
\begin{equation}\label{eq:signal}
pp \to t\bar t \to W^\mp bH^\pm b,\qquad H^\pm \to t^*b \to W^\pm b\bar b\,,
\end{equation}
with a corresponding production cross section of
\begin{equation}\begin{split}
  & 2\, \sigma_{t\bar t} \times \left(1- {\rm Br}(t\to H^\pm b)\right) \\
  & \hspace{2.5cm}\times {\rm Br}(t\to H^\pm b) \times {\rm Br}(H^\pm \to t^* b)\,.
\end{split}\end{equation}
Here, the factor of two accounts for the possibility that either the top or antitop decays via the charged-Higgs channel, and the top-quark pair production cross section is $\sigma_{t\bar t} \equiv \sigma(pp\to t\bar t)= 832\pm 49$\,pb for $m_t = 172.5$\,GeV at the 13\,TeV LHC, after including corrections at the next-to-next-to-leading order (NNLO) in QCD~\cite{Czakon:2011xx}. 

We reinterpret the recent ATLAS measurement of fiducial cross sections for dileptonic $t\bar t$ production with additional $b$-jets at $\sqrt{s}=13$~TeV, based on 140~fb$^{-1}$ of data~\cite{ATLAS:2024aht}, to constrain the charged-Higgs signal defined in Eq.~\eqref{eq:signal}. To validate our setup, we generate SM events for the processes $pp \to t\bar t$ and $pp\to t\bar tb\bar b$ with {\tt MadGraph5\_aMC\_v3.5.8}~\cite{Alwall:2014hca, Frederix:2018nkq}, convoluting  matrix elements evaluated at the next-to-leading order (NLO) accuracy in QCD with the NLO set of {\tt NNPDF3.0} parton distribution functions~\cite{NNPDF:2017mvq, Buckley:2014ana} and relying on {\tt MadSpin}~\cite{Artoisenet:2012st} and {\tt MadWidth}~\cite{Alwall:2014bza} for the treatment of heavy-particle decays. The obtained hard-scattering events are then interfaced with {\tt Pythia 8.3}~\cite{Bierlich:2022pfr} using the {\tt A14} tune~\cite{TheATLAScollaboration:2014rfk} to model parton showering and hadronization effects. By contrast, the charged Higgs signal is generated at leading order (LO) on the basis of the 2HDM UFO libraries generated within the \textsc{FeynRules}/UFO framework~\cite{Christensen:2009jx, Alloul:2013bka, Degrande:2011ua, Darme:2023jdn}, fixing the charged-Higgs mass in the 100 -- 165\,GeV range.

For the reconstruction and the selection of leptons and jets, we closely follow the ATLAS analysis of Ref.~\cite{ATLAS:2024aht}. Jets are clustered with the anti-$k_T$ algorithm~\cite{Cacciari:2008gp} as implemented in {\tt FastJet 3.3.4}~\cite{Cacciari:2011ma}, using a radius parameter $R=0.4$. Reconstructed events are then required to contain an opposite-sign $e^\pm\mu^\mp$ pair originating from the decay of the two final-state $W$ bosons. We next apply the remaining analysis-specific requirements on the lepton and jet multiplicities, together with the corresponding kinematic object selections, so that the selected event samples match the fiducial signal regions defined in the ATLAS analysis. Four fiducial signal regions are considered, based on the multiplicity of identified $b$-tagged, $c$-tagged and light jets $j$,
\begin{align}
  &\geq 3b,\quad 
    \geq 3b\geq1c/j,\quad
    \geq 4b,\quad
    \geq 4b\geq1c/j.
\end{align}

\begin{figure}
  \centering
  \includegraphics[width=0.48\textwidth]{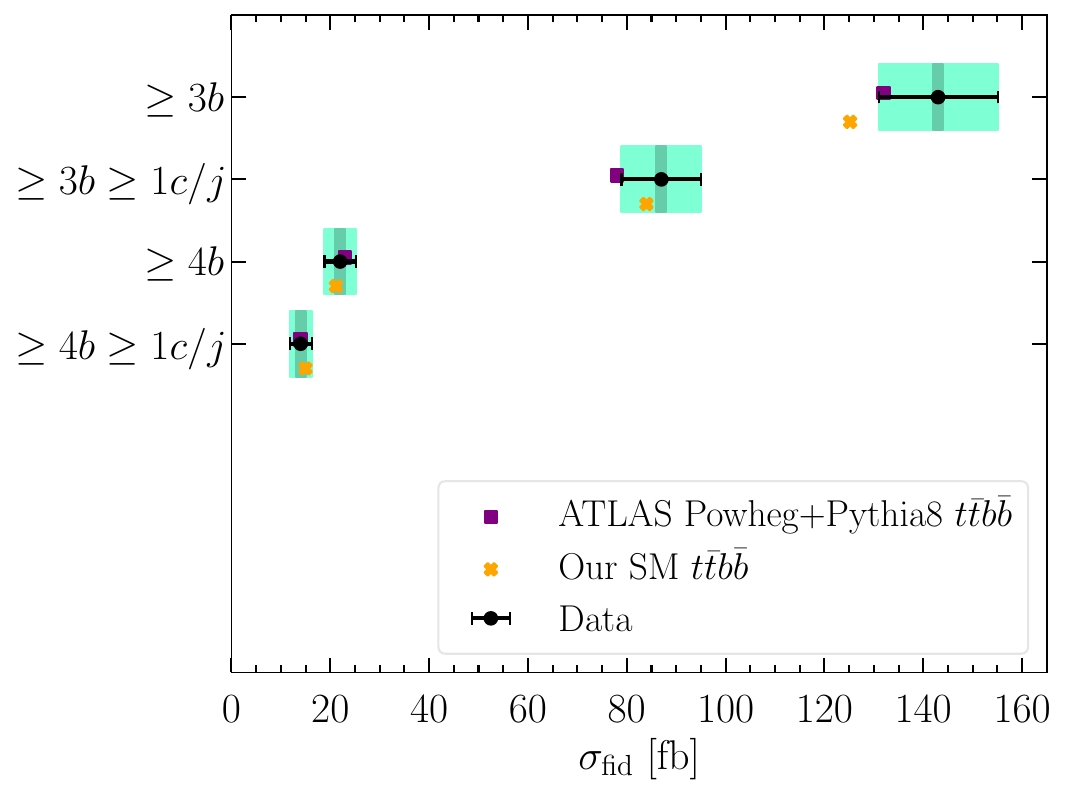}
  \caption{Measured fiducial cross sections in the four considered ATLAS signal regions compared with the ATLAS \textsc{Powheg+Pythia8} predictions, and our simulated SM \(t\bar tb\bar b\) predictions. The inner dark band denotes the statistical uncertainty of the data, while the outer light band includes the total uncertainty from statistical, instrumental, and theoretical sources.}
  \label{fig:fiducial_xs}
\end{figure}

To validate our implementation, we compare in Figure~\ref{fig:fiducial_xs} the measured fiducial cross sections in the four ATLAS signal regions with the ATLAS \textsc{Powheg+Pythia8} predictions~\cite{Jezo:2018yaf} and with our simulated SM $t\bar t b\bar b$ sample. The four regions probe different heavy-flavor topologies: the inclusive $\geq3b$ categories have the largest fiducial rates and therefore benefit from smaller statistical uncertainties, whereas the $\geq4b$ categories are more directly enriched in events with two additional $b$-jets, but suffer from a smaller event yield. Our simulation reproduces the hierarchy of the fiducial cross sections across the four regions and remains compatible with the measured values within the experimental uncertainties. This agreement validates the object selection, flavor categorization and fiducial-region implementation used in the reinterpretation below.

\section{Limits on light charged Higgs bosons from \texorpdfstring{$t\bar tb\bar b$}{ttbb} measurements} \label{sec:limits}

\subsection{Model-independent fiducial limits}

We derive model-independent constraints by comparing the measured $t\bar t b\bar b$ fiducial cross sections by the ATLAS collaboration with the corresponding SM theoretical predictions supplemented by the charged-Higgs contribution. In doing so, we neglect the interference between the SM and the charged-Higgs amplitudes. This differs from the familiar case of heavy-Higgs interference in $gg\to t\bar t$~\cite{Gaemers:1984sj, Dicus:1987fk, Dicus:1994bm, Bernreuther:1997gs, Bahl:2025you, Flacke:2025dwk}. In the present case, the SM and beyond the SM (BSM) contributions do not share the same resonant structure. The charged-Higgs signal contains an intermediate on-shell $H^\pm$ state produced in the rare decay $t\to H^\pm b$, whereas the SM $t\bar t b\bar b$ amplitude is smooth in the corresponding virtuality. In addition, the kinematics of the $Wb$ systems are different. In SM $t\bar t b\bar b$ production, both $Wb$ systems originate from nearly on-shell top quarks, while in the signal one of them arises from the off-shell cascade $H^\pm\to t^*b\to W^\pm b\bar b$. Together with the narrow charged-Higgs width, this resonant and kinematic mismatch suppresses the interference after integration over the fiducial phase space.

The sensitivity of the resulting bounds is limited by the sizable theoretical uncertainty of the SM $t\bar t b\bar b$ prediction, which is approximately 30\% in the relevant fiducial regions~\cite{Bevilacqua:2021cit, Bevilacqua:2022twl}. We treat this uncertainty using the $R$-fit prescription~\cite{Hocker:2001xe}. In practice, for each charged-Higgs signal hypothesis, the SM prediction is allowed to vary anywhere within its quoted theoretical uncertainty band. We then choose the value in this band that gives the best agreement with the measured fiducial cross section after the signal contribution has been added. If the measurement can still be reproduced in this way, the corresponding signal hypothesis is not penalized by the theory uncertainty. A contribution to the test statistic arises only from the residual difference that cannot be absorbed by varying the SM prediction within its allowed range. In addition, the experimental uncertainty, dominated by systematic effects, is treated as Gaussian. The resulting model-independent limits on an additional non-SM contribution to each fiducial signal region are summarized in Table~\ref{tab:fiducial_limits}.

\begin{table}
  \setlength{\tabcolsep}{12pt} \renewcommand{\arraystretch}{1.2}
  \centering
  \begin{tabular}{lcc}
    \toprule
    Signal region & 68\% CL & 95\% CL \\
    \midrule
    $e\mu+\geq 3b$ & 56\,fb & 70\,fb \\
    $e\mu+\geq 3b\geq 1c/j$ & 36\,fb & 46\,fb \\
    $e\mu+\geq 4b$ & 7\,fb & 11\,fb \\
    $e\mu+\geq 4b\geq 1c/j$ & 5\,fb & 8\,fb \\
    \bottomrule
  \end{tabular}
  \caption{Observed 68\% and 95\% CL upper limits on an additional non-SM contribution to the fiducial cross section in each ATLAS signal region treated independently.}
  \label{tab:fiducial_limits}
\end{table}

The four fiducial regions are inclusive, since they impose only a lower limit on the jet multiplicities, and they therefore partially overlap. Although the corresponding correlations could in principle be estimated from simulations, we derive the constraints separately in each region. Given the sizable theoretical uncertainty affecting the SM $t\bar t b\bar b$ prediction, a statistical combination of the regions is not expected to improve the sensitivity significantly. Furthermore, we have also checked that the relevant differential distributions have very similar shapes for the SM and the charged-Higgs contributions. Including this shape information is therefore expected to have only a marginal impact on the bounds compared with an analysis based solely on integrated fiducial cross sections.

\begin{figure}
  \centering
  \includegraphics[width=0.49\textwidth]{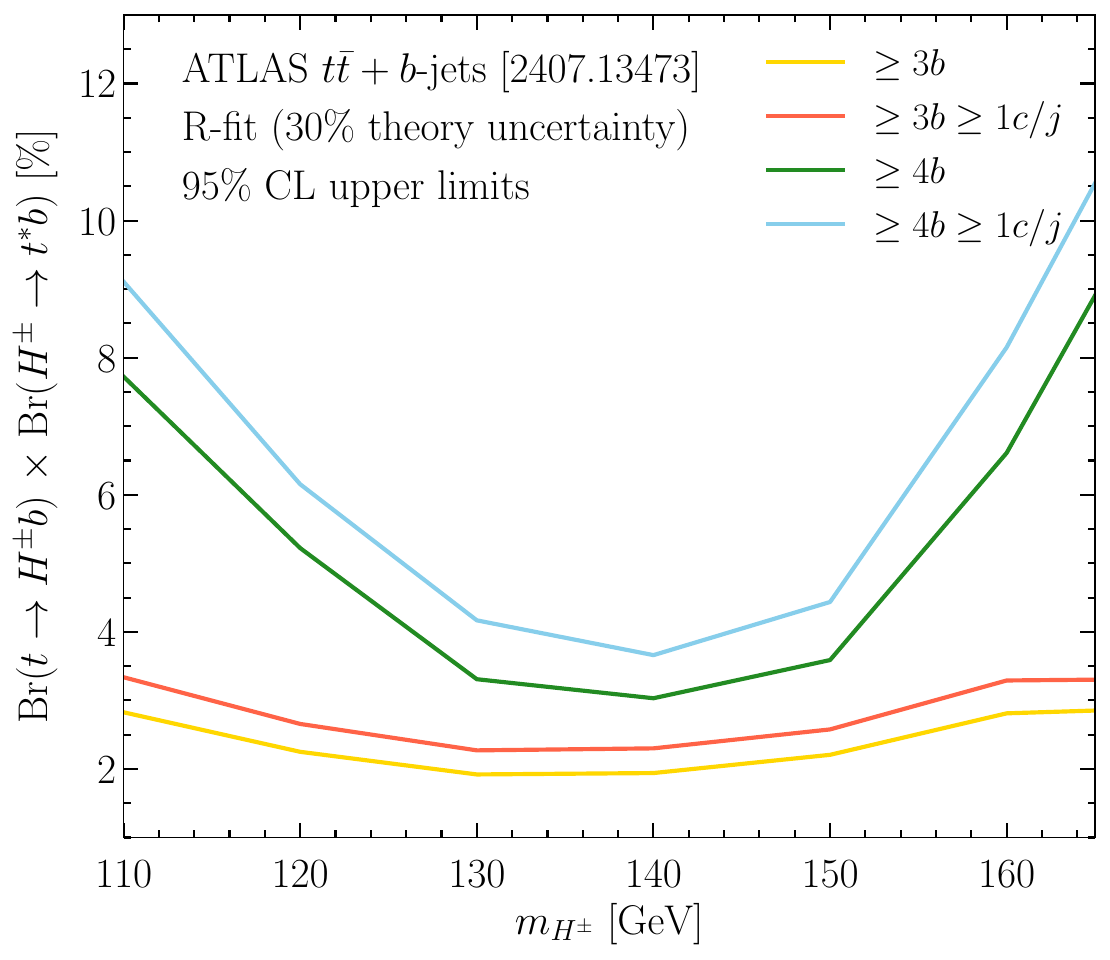}
  \caption{Observed 95\% CL upper limits on the product  ${\rm Br}(t\to H^\pm b)\times{\rm Br}(H^\pm\to t^*b)$ as a function of the charged-Higgs mass, obtained separately in the four ATLAS fiducial signal regions using an $R$-fit treatment of the SM theory uncertainty.}
  \label{fig:limit_mu}
\end{figure}

Figure~\ref{fig:limit_mu} shows the resulting 95\% confidence-level (CL) upper limits on the product of branching ratios ${\rm Br}(t\to H^\pm b)\times{\rm Br}(H^\pm\to t^*b)$ as a function of $m_{H^\pm}$ for all four signal regions. The mass dependence is governed mainly by the competing kinematics of the two stages of the decay chain. For a low charged-Higgs mass, the decay products of the off-shell top quark are typically soft and are therefore less likely to satisfy the jet multiplicity requirements. As the value of $m_{H^\pm}$ increases, these decay products become harder and the acceptance improves, yielding the strongest sensitivity around $m_{H^\pm}\simeq130$--$150$\,GeV. Closer to the top-quark threshold, however, the bottom quark produced in the $t\to H^\pm b$ decay becomes increasingly soft because of the shrinking phase space, causing the sensitivity to deteriorate again. Among the four fiducial regions, the inclusive $\geq3b$ category provides the strongest observed limit over the full mass range, with the $\geq3b\geq1c/j$ region giving a comparable but slightly weaker constraint. The $\geq4b$ and $\geq4b\geq1c/j$ categories are less sensitive despite being more enriched in events with additional bottom quarks because of the reduced efficiency for reconstructing and tagging all the $b$-jets in the signal events.

Over a charged-Higgs mass range of $110$--$165$\,GeV, the inclusive $\geq3b$ region yields observed 95\% CL upper limits of 1.9\%--2.9\% on the product ${\rm Br}(t\to H^\pm b)\times {\rm Br}(H^\pm \to t^* b)$. This shows that despite the sizable theoretical uncertainty affecting the SM $t\bar t b\bar b$ prediction, existing fiducial measurements already provide a meaningful sensitivity to a charged-Higgs decay mode that has not been targeted by dedicated searches.

The limits derived above depend only on the assumed production and decay topology, and are otherwise independent of the structure of the scalar sector. They can therefore be directly interpreted in models featuring a light charged Higgs boson with a sizable branching fraction into $t^*b$. We now turn to such model-dependent interpretations.

\subsection{Interpretation in extended Higgs sectors}
\label{sec:interpreation}

In this section, we translate the model-independent limits on the product ${\rm Br}(t\to H^\pm b)\times {\rm Br}(H^\pm\to t^*b)$ into constraints on representative models with extended Higgs sectors. We parameterize the relevant charged-Higgs interactions as
\begin{equation}\label{eq:lag}\begin{split}
  \mathcal{L} = &\ 
    \frac{\sqrt{2}}{v} \big[ \bar u_i V_{ij} \left(\bar m_{u_i}\xi_{u_i} P_L + \bar m_{d_j}\xi_{d_j} P_R \right) d_j H^+ \\
    & \qquad + \bar m_{\ell_i}\xi_{\ell_i} \bar\nu_{\ell_i} P_R \ell_i \big] + \mathrm{H.c.},
\end{split}\end{equation}
where $u_i,d_j,\ell_i$ denote the up-type quarks, down-type quarks, and charged leptons, respectively, $v$ is the electroweak vacuum expectation value, $P_{L,R}$ are the left-handed and right-handed chirality projectors, and $V_{ij}$ the relevant CKM matrix elements. The parameters $\xi_{u,d,\ell}$ are vectors in the flavor space and encode the model-dependent Yukawa structure. 

The partial width for the $t\to H^\pm b$ decay is given by
\begin{equation}\label{eq:LO_Hp_width} \begin{split}
  &\Gamma(t\to H^\pm b) = 
    \frac{G_F|V_{tb}|^2}{8\sqrt{2}\pi m_t^3} \lambda^{\frac{1}{2}}(m_t^2, m_{H^\pm}^2, m_b^2) \\
    &\ \times \bigg[ \Big(\bar m_t^2 |\xi_t|^2 + \bar m_b^2 |\xi_b|^2\Big) \Big(m_t^2 + m_b^2 - m_{H^\pm}^2\Big) \\
    &\hspace{2cm} + 4 m_t m_b \bar m_t \bar m_b \, \Re\big\{\xi_t \xi_b^*\big\} \bigg], 
\end{split}\end{equation}
where $G_F$ is the Fermi coupling, and $\lambda(a,b,c)$ is the K\"all\'en function. The masses $m_t$ and $m_b$ within the phase-space factors are taken as pole masses, whereas $\bar m_t$ and $\bar m_b$ denote the corresponding $\overline{\rm MS}$ running masses determining the Yukawa couplings evaluated at the scale $m_t$. We use~\cite{ParticleDataGroup:2024cfk, Huang:2020rtx, Aparisi:2021tym, Ma:2024xeq}
\begin{equation} \begin{split}
  m_t=172.5\,\mathrm{GeV}, \qquad 
  &m_b=5.37\,\mathrm{GeV}, \\
  \bar m_t(m_t)=162.7\,\mathrm{GeV},\qquad
  &\bar m_b(m_t)=2.6\,\mathrm{GeV}.
\end{split}\end{equation}
The LO expression of Eq.~\eqref{eq:LO_Hp_width} is then corrected for higher-order QCD effects using the NNLO results of Ref.~\cite{Shen:2022yuo}. For the central renormalization scale $\mu_R=m_t$, these corrections increase the partial width by about 1\%, 3\%, and 5\% at $m_{H^\pm}=120$, 140, and 150\,GeV, respectively, with the enhancement thus growing as the charged-Higgs mass approaches the kinematic threshold. The corresponding branching fraction is 
\begin{align}
{\rm Br}(t\to H^\pm b) = \frac{\Gamma(t\to H^\pm b)}{\Gamma(t\to H^\pm b) + \Gamma (t)_{\rm SM}},
\end{align}
where the SM top-quark width $\Gamma (t)_{\rm SM}= 1.326$\,GeV includes all known higher-order QCD and electroweak corrections, as well as finite $b$-quark mass effects~\cite{Chen:2022wit, ParticleDataGroup:2024cfk}.

The charged-Higgs branching fraction relevant for our analysis is defined as
\begin{align*}
{\rm Br}(H^\pm\to t^*b) = \frac{\Gamma(H^\pm\to t^*b \to W^\pm b\bar b)}{\Gamma({H^\pm})}.
\end{align*}
As the QCD corrections to the three-body decay $H^\pm\to t^*b \to W^\pm b\bar b$ are not known, we evaluate this partial width at leading order. Furthermore, in the scenarios considered below, we assume that other additional neutral scalars are heavy enough that they do not open extra decay modes of the light charged Higgs boson. The total width $\Gamma(H^\pm)$ is then determined by the fermionic channels only, including the off-shell $t^*b$ mode as well as the conventional light-quark and leptonic modes when allowed by the Yukawa structure. For completeness, the relevant partial-width expressions are collected in Appendix~\ref{sec:app}.

Before turning to individual benchmark models, let us comment on the applicable highly model-dependent flavor constraints, which depend strongly on the charged-Higgs Yukawa structure. In the minimal Type-II and Type-Y 2HDMs, the light charged-Higgs region of the parameter space is excluded by the $\bar B\to X_s\gamma$ bound, which implies $m_{H^\pm}\gtrsim 570$\,GeV~\cite{Misiak:2017bgg}. In Type-I and Type-X models where the quark couplings scale as $\cot\beta$ with $\tan\beta\equiv v_2/v_1$ denoting the ratio of the two Higgs-doublet vacuum expectation values, flavor and electroweak observables mainly constrain the low-$\tan\beta$ region, with the most relevant bounds arising from $\bar B\to X_s\gamma$, $B_{d,s}$ mixing, and $Z\to b\bar b$~\cite{Jung:2010ik, Enomoto:2015wbn, Coutinho:2024zyp}. For the up-type aligned, top-philic, and top-specific benchmarks examined below, the absence or suppression of down-type and leptonic charged-Higgs Yukawa couplings weakens these bounds. Moreover, unlike direct searches for the charged Higgs boson itself, indirect flavor limits can in principle be modified by the interference with additional new-physics contributions. We therefore present the collider reinterpretation as a direct constraint on the charged-Higgs topology of Eq.~\eqref{eq:signal}, and discuss its impact in each benchmark model separately.

First, we consider the top-philic limit of the general Type-III 2HDM~\cite{Crivellin:2013wna}. In the Higgs basis, this corresponds to a Yukawa structure in which only the top-quark entry of the up-type coupling is nonzero, while the down-type and charged-lepton couplings vanish,
\begin{equation}\label{eq:topphilic}
    \xi_u=(0,0,\xi_t),\qquad
    \xi_d = \xi_\ell = (0,0,0).
\end{equation}
The charged Higgs therefore couples to the top quark, but not to the light up-type quarks, down-type quarks, or charged leptons. Consequently, the conventional light charged-Higgs decay modes $H^\pm\to cs$, $H^\pm\to cb$, and $H^\pm\to\tau\nu$ are absent, and the limits originating from the searches targeting these final states do not apply. With the additional scalar states assumed to be heavy, the total charged-Higgs width is then dominated by the off-shell decay $H^\pm\to t^*b\to W^\pm b\bar b$, so that the branching ratio ${\rm Br}(H^\pm\to t^*b)$ is close to unity. The signal rate is therefore controlled mainly by the rare top decay $t\to H^\pm b$.

\begin{figure}
  \centering
  \includegraphics[width=0.49\textwidth]{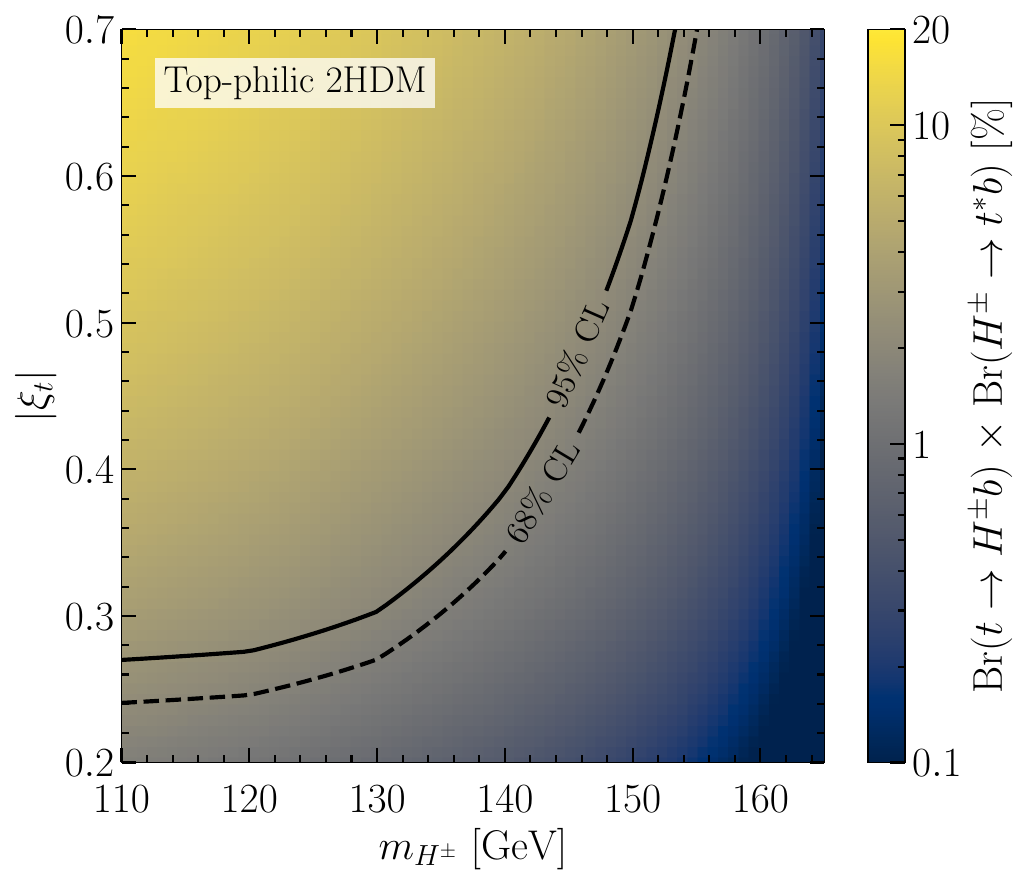}
  \caption{Product of branching ratios ${\rm Br}(t\to H^\pm b)\times{\rm Br}(H^\pm\to t^*b)$ in the top-philic 2HDM, shown in the $(m_{H^\pm},|\xi_t|)$ plane. The solid and dashed black curves denote the observed 95\% and 68\% CL limits from the $t\bar tb\bar b$ reinterpretation, respectively, with the region above each contour being excluded.}
\label{fig:limit_top}
\end{figure}

Figure~\ref{fig:limit_top} shows the resulting product ${\rm Br}(t\to H^\pm b)\times{\rm Br}(H^\pm\to t^*b)$ in the $(m_{H^\pm},|\xi_t|)$ plane, together with the 68\% and 95\% CL exclusion contours obtained from the reinterpretation of the $t\bar t b\bar b$ measurements. In the top-philic benchmark model, the expression for $\Gamma(t\to H^\pm b)$ reduces to a term proportional to $|\xi_t|^2$, up to the charged-Higgs mass dependence induced by the phase space and by the higher-order QCD corrections discussed above. The color gradient in Figure~\ref{fig:limit_top} therefore mostly reflects the growth of the rare top-decay branching ratio with $|\xi_t|$ and its suppression as $m_{H^\pm}$ approaches the kinematic threshold. The observed exclusion reaches down to approximately $|\xi_t|\simeq0.27$ in the most sensitive mass range. Although the model-independent bound on the branching-ratio product is strongest around $m_{H^\pm}\simeq130$--$150$\,GeV, its translation into a bound on $|\xi_t|$ is also controlled by the phase-space suppression of the rare top decay $t\to H^\pm b$. As $m_{H^\pm}$ approaches the kinematic threshold, this suppression rapidly reduces ${\rm Br}(t\to H^\pm b)$ for a fixed value of $|\xi_t|$, so that larger couplings are required to produce the same signal rate and the exclusion contour moves upward.

We next relax the purely top-philic structure of Eq.~\eqref{eq:topphilic} and consider the up-type aligned limit of the 2HDM. In this case, the charged-Higgs couplings to all up-type quarks are controlled by a common parameter, while the down-type and charged-lepton couplings vanish,
\begin{equation}
  \xi_u=\xi_u(1,1,1),\qquad
  \xi_d=\xi_\ell=(0,0,0).
\end{equation}
Compared with the top-philic scenario, the nonzero charm coupling opens two-body decays of the form $H^\pm\to c d_j$ and $H^\pm\to u d_j$. Among these, the $cs$ channel dominates because it is controlled by the charm Yukawa coupling and the CKM-favored matrix element $V_{cs}$. The $cd$ mode is CKM-suppressed at the few-percent level, the $cb$ one is further suppressed, and the modes involving an up quark are negligible because of the small up-quark mass. The charged-Higgs branching pattern is therefore mainly governed by the competition between the two-body decay $H^\pm\to cs$ and the off-shell mode $H^\pm\to t^*b$. The former benefits from favorable two-body phase space and dominates at lower charged-Higgs masses, whereas the latter is driven by the top Yukawa coupling and becomes increasingly important as $m_{H^\pm}$ approaches the top-quark threshold. 

\begin{figure}
  \centering
  \includegraphics[width=0.49\textwidth]{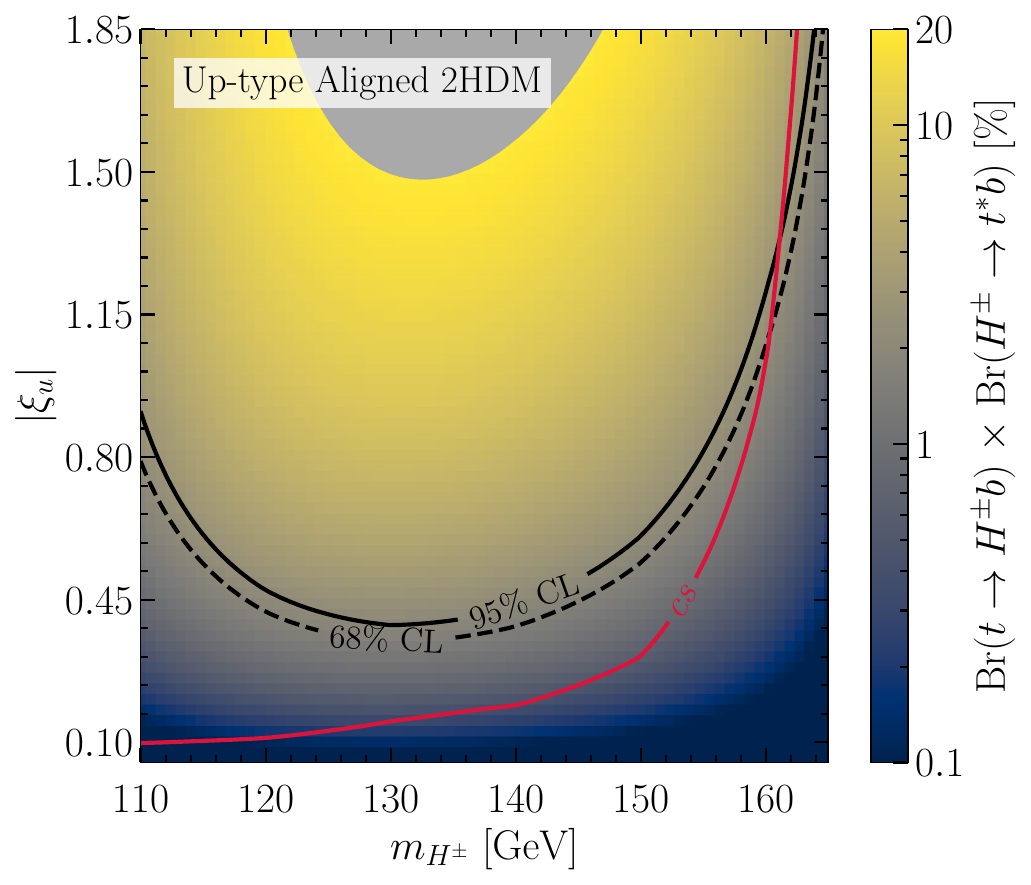}
  \caption{Product of branching ratios ${\rm Br}(t\to H^\pm b)\times{\rm Br}(H^\pm\to t^*b)$ in the up-type aligned limit, shown in the $(m_{H^\pm},|\xi_u|)$ plane. The solid and dashed black curves show the observed 95\% and 68\% CL exclusions from the $t\bar t b\bar b$ reinterpretation, respectively, with the regions above the contours being excluded. The red curve indicates the observed 95\% CL exclusion from the dedicated $H^\pm\to cs$ search~\cite{ATLAS:2024oqu}, and the gray region corresponds to ${\rm Br}(t\to H^\pm b)>20\%$.}
  \label{fig:limit_up}
\end{figure}

Figure~\ref{fig:limit_up} shows the product ${\rm Br}(t\to H^\pm b)\times {\rm Br}(H^\pm\to t^*b)$ in the $(m_{H^\pm},|\xi_u|)$ plane, together with the exclusion contours obtained from the $t\bar t b\bar b$ reinterpretation. Since the same coupling $\xi_u$ controls both the top and charm charged-Higgs interactions, the production rate through $t\to H^\pm b$ grows with $|\xi_u|^2$, while the relative importance of the $cs$ and $t^*b$ decay modes is governed mainly by the charged-Higgs mass. At lower charged-Higgs masses, the two-body charm modes remain important and dilute the branching fraction into $t^*b$. As $m_{H^\pm}$ increases, the off-shell top contribution is enhanced by the large top Yukawa coupling and by the increasing phase space available in $H^\pm\to t^*b$, making the $t^*b$ mode progressively more relevant. 

For comparison, the red curve shows the 95\% CL exclusion from the dedicated $H^\pm\to cs$ search, obtained by confronting the predicted ${\rm Br}(t\to H^\pm b)\times{\rm Br}(H^\pm\to cs)$ rate with the experimental upper limit of Ref.~\cite{ATLAS:2024oqu}. This search provides the stronger constraint over most of the charged-Higgs mass range, excluding substantially smaller values of $|\xi_u|$ than the $t\bar t b\bar b$ reinterpretation. Its sensitivity, however, weakens as $m_{H^\pm}$ approaches the top-quark mass, because the branching fraction into the $t^*b$ final state increases at the expense of the charm mode. As a result, the two constraints become comparable around $m_{H^\pm}\simeq160\,{\rm GeV}$, with the $t\bar t b\bar b$ reinterpretation becoming slightly stronger closer to the kinematic endpoint.

At intermediate charged-Higgs masses and large $|\xi_u|$, the branching fraction for the rare top decay can become very large. The gray region included in the plot corresponds to ${\rm Br}(t\to H^\pm b)>20\%$. Such a large nonstandard top-quark branching fraction would significantly affect the total top width and is in tension with the precise indirect CMS determination of this quantity, subject to the assumptions entering that extraction~\cite{CMS:2014mxl}. We therefore do not interpret this region quantitatively.

We continue our analysis with the top-specific 2HDM, which provides a Yukawa structure intermediate between the top-philic scenario and the canonical $Z_2$-symmetric ones briefly addressed below. Here, one scalar doublet couples exclusively to the top quark, while the other couples to all remaining fermions. In the convention adopted in Eq.~\eqref{eq:lag}, the charged-Higgs couplings are given by
\begin{equation}\begin{split}
  &\xi_u = (-\tan\beta, -\tan\beta, \cot\beta),\\
  &\xi_d = \xi_\ell = (\tan\beta, \tan\beta, \tan\beta).
\end{split}\end{equation}
The charged-Higgs phenomenology is therefore controlled by two competing regimes. At low $\tan\beta$, the top coupling is enhanced, which increases both the rare top decay rate $t\to H^\pm b$ and the partial width for the off-shell charged-Higgs decay $H^\pm\to t^*b$. At the same time, the couplings responsible for the conventional $cs$, $cb$, and $\tau\nu$ modes are suppressed. Conversely, at larger $\tan\beta$, the light-quark, bottom-quark, and leptonic charged-Higgs couplings grow, while the impact of the top contribution decreases. 

\begin{figure}
  \centering
  \includegraphics[width=0.49\textwidth]{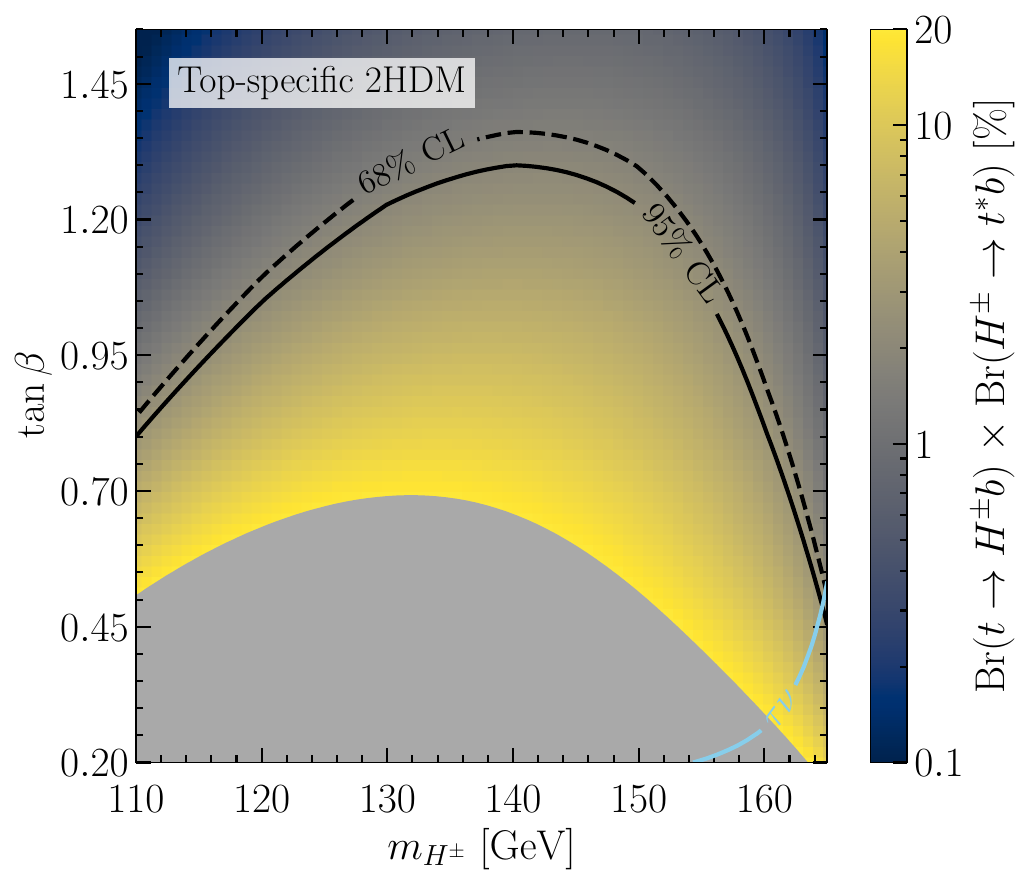}
  \caption{Product of branching ratios ${\rm Br}(t\to H^\pm b)\times{\rm Br}(H^\pm\to t^*b)$ in the top-specific 2HDM, shown in the $(m_{H^\pm},\tan\beta)$ plane. The solid and dashed black curves denote the observed 95\% and 68\% CL exclusions from the $t\bar t b\bar bb$ reinterpretation, respectively, while the blue curve shows the observed 95\% CL exclusion from dedicated $H^\pm\to\tau\nu$ searches~\cite{CMS:2019bfg, ATLAS:2024hya}. The region below the black contour is excluded by the $t\bar t b\bar b$ reinterpretation, while the region above the blue contour is excluded by the $\tau\nu$ searches. The gray region corresponds to ${\rm Br}(t\to H^\pm b)>20\%$.}
\label{fig:limit_topspecific}
\end{figure}

Figure~\ref{fig:limit_topspecific} shows the product ${\rm Br}(t\to H^\pm b)\times{\rm Br}(H^\pm\to t^*b)$ in the $(m_{H^\pm},\tan\beta)$ plane, together with the corresponding exclusion contours. The colour gradient reflects the above-mentioned interplay between the enhanced top coupling at low $\tan\beta$ and the growing importance of the conventional charged-Higgs decay modes at larger $\tan\beta$. For small $\tan\beta$, the coupling $\xi_t=\cot\beta$ enhances the relevant production rate through the larger partial width for the $t\to H^\pm b$ decay, and additionally favors the $H^\pm\to t^*b$ channel. The sensitivity is strongest near the upper part of the charged-Higgs mass range where the off-shell top mode is most relevant, although the phase-space suppression of the $t\to H^\pm b$ decay still limits the reach close to the kinematic endpoint. At very low $\tan\beta$, the rare top-decay branching fraction becomes large enough to enter the gray region featuring significant nonstandard top-quark branching fraction, which we do not interpret quantitatively. The relevant exclusion from the $t\bar t b\bar b$ reinterpretation therefore appears as a band just above this region.

The blue contour comparatively shows the 95\% CL exclusion originating from dedicated $H^\pm\to\tau\nu$ searches~\cite{CMS:2019bfg, ATLAS:2024hya}, which provide the strongest conventional constraint in the parameter space region considered. These searches are highly efficient at moderate and large $\tan\beta$, where the charged-Higgs coupling to leptons is enhanced. By contrast, their sensitivity is reduced at very low $\tan\beta$, because the leptonic branching fraction becomes suppressed. In this regime the enhanced top coupling instead makes the $t^*b$ channel the relevant probe, so that our results highlight how the $t\bar t b\bar b$ measurements can exclude a low-$\tan\beta$ corner of the parameter space that is not covered by conventional light charged-Higgs searches.

Finally, we briefly comment on the canonical $Z_2$-symmetric 2HDMs, namely the Type-I, Type-X (lepton-specific), and Type-Y (flipped) models. In the convention adopted above, their charged-Higgs Yukawa couplings follow the patterns shown in Table~\ref{tab:canonical}. In all three cases, the rare top decay $t\to H^\pm b$ is enhanced at low $\tan\beta$, where the top-mass contribution to the charged-Higgs coupling scales as $\cot\beta$. For fixed $\tan\beta$, the corresponding branching fraction decreases as $m_{H^\pm}$ approaches the kinematic endpoint of the top decay. These canonical models, however, also predict conventional light charged-Higgs decay modes that are efficiently covered by dedicated searches. In the Type-I 2HDM, the partial widths for the $cs$, $cb$, and $\tau\nu$ channels all scale with $\cot\beta$, while in Type-X the leptonic mode is enhanced at large $\tan\beta$ and in Type-Y the down-type quark modes are enhanced instead. As a result, the existing searches for the $H^\pm\to cs$, $cb$, and $\tau\nu$ decays already provide stronger constraints in the relevant regions of the parameter space~\cite{CMS:2019bfg, CMS:2020osd, ATLAS:2023bzb, ATLAS:2024oqu, ATLAS:2024hya}. The $t\bar t b\bar b$ reinterpretation is instead most relevant in scenarios with nonstandard Yukawa structures, such as the top-philic, up-type aligned, and top-specific scenarios, where the conventional charged-Higgs decay modes are absent or suppressed.

\begin{table}
  \setlength{\tabcolsep}{16pt} \renewcommand{\arraystretch}{1.3}
  \centering
  \begin{tabular}{cccc}
    \toprule
    Model & $\xi_u$ & $\xi_d$ & $\xi_\ell$ \\
    \midrule
    Type-I & $\cot\beta$ & $-\cot\beta$ & $-\cot\beta$ \\
    Type-X & $\cot\beta$ & $-\cot\beta$ & $\tan\beta$ \\
    Type-Y & $\cot\beta$ & $\tan\beta$ & $-\cot\beta$ \\
    \bottomrule
  \end{tabular}
  \caption{Charged-Higgs Yukawa coupling modifiers in the canonical $Z_2$-symmetric 2HDMs considered.}
  \label{tab:canonical}
\end{table}

\section{Summary and prospects}\label{sec:prospects}

We have investigated the phenomenology of light charged Higgs bosons produced in the rare top-quark decay $t\to H^\pm b$, and subsequently decaying through the previously unexplored mode $H^\pm\to t^*b\to W^\pm b\bar b$ with an off-shell top quark. This decay chain gives rise to a $t\bar t b\bar b$-like final state, and can therefore be probed by fiducial cross-section measurements of $t\bar t$ production in association with additional $b$-jets. By reinterpreting recent ATLAS measurements of dileptonic $t\bar t b\bar b$ production, we have derived 95\% CL upper limits of $1.9\%$--$2.9\%$ on the product of branching ratios ${\rm Br}(t\to H^\pm b)\times{\rm Br}(H^\pm\to t^*b)$ for charged-Higgs masses between 110 and 165\,GeV. The sensitivity is strongest at intermediate masses, where the decay products of the off-shell top quark are sufficiently energetic to satisfy the fiducial jet multiplicity requirements while the rare top decay $t\to H^\pm b$ is not yet strongly phase-space suppressed.

We have interpreted these limits in several representative 2HDM scenarios. In the canonical $Z_2$-symmetric Type-I, Type-X, and Type-Y models, the parameter space regions probed by the $t\bar t b\bar b$ channel are already more strongly constrained by searches for the conventional $cs$, $cb$, and $\tau\nu$ decay modes. In the up-type aligned limit, the $H^\pm\to cs$ search remains dominant over most of the mass range, but the $t\bar t b\bar b$ reinterpretation becomes competitive close to the top-quark threshold. In the top-philic benchmark model where the nonstandard charged-Higgs interaction is restricted to the top-quark sector, conventional light-quark and leptonic searches do not apply, and the present reinterpretation provides the relevant direct constraint. Finally, in the top-specific 2HDM, the $t^*b$ and $\tau\nu$ searches probe complementary regions: the $\tau\nu$ search dominates over most of the parameter space, whereas the $t\bar t b\bar b$ measurements exclude a low-$\tan\beta$ corner near the upper end of the charged-Higgs mass range that is not covered by conventional light charged-Higgs searches.

Our results demonstrate that recasts of existing SM measurements can already provide sensitivity to nonstandard top decays that have not yet been targeted directly. They motivate dedicated searches by the ATLAS and CMS collaborations for the $t\to H^\pm b$ with $H^\pm\to t^*b$ channel, exploiting the characteristic enhancement in the $b$-jet multiplicity and the kinematics of the off-shell top decay. Such analyses could substantially improve upon the present reinterpretation, especially through the use of differential distributions, optimized multivariate discriminants, and the increased statistics of the LHC Run~3 and high-luminosity phase. Furthermore, improved SM predictions for $t\bar t b\bar b$ production, together with a more complete treatment of the correlations among the fiducial regions, would reduce the dominant uncertainty and significantly sharpen the sensitivity to new-physics effects.

\acknowledgments SA is supported by the Deutsche Forschungsgemeinschaft (DFG, German Research Foundation) under grant 396021762 -- TRR 257. GC acknowledges financial support from the NextGenerationEU funds of the MUR, through the PRIN2022 grant Nr.~2022EZ3S3F.

\appendix
\section{Charged-Higgs partial widths}
\label{sec:app}

For completeness, we collect the fermionic charged-Higgs partial widths used in our numerical analysis. In the benchmark scenarios considered in this work, bosonic charged-Higgs decays are either kinematically closed or absent, and in the following we denote the charged-Higgs mass by $M\equiv m_{H^\pm}$. With $u,d,\ell$ denoting generic up-type quarks, down-type quarks, and charged leptons, the two-body fermionic widths read
\begin{equation} \begin{split}
  & \Gamma (H^\pm\to u\bar d) = \frac{3G_F|V_{ud}|^2}{4\sqrt{2}\pi M^3} \, \lambda^{\frac{1}{2}}(M^2, m_u^2, m_d^2)\\
   & \qquad \times \bigg[ \Big(\bar m_u^2 |\xi_u|^2 + \bar m_d^2 |\xi_d|^2\Big) \Big(M^2 - m_u^2 - m_d^2\Big) \\
   & \hspace{2cm} - 4\, m_u\, m_d\, \bar m_u\, \bar m_d\, \,\Re\big\{\xi_u \xi_d^*\} \bigg], \\
  & \Gamma(H^\pm\to \ell^\pm\nu) = \frac{G_F}{4\sqrt{2}\pi M^3}\, \bar m_\ell^2 |\xi_\ell|^2 \left(M^2 - m_\ell^2\right)^2.\\
\end{split}\end{equation}
Here, $\bar m_f$ denotes the running fermion mass entering the Yukawa coupling, while $m_f$ stands for the pole mass used in the phase space and $\lambda(x,y,z)$ stands for the conventional K\"all\'en function.

The three-body width relevant for the off-shell top decay mode is written as
\begin{align}
\Gamma&(H^\pm\to t^* b \rightarrow W^\pm b \bar{b}) =\int_{s_2^-}^{s_2^+} {\rm d} s_2  \int_{s_1^-}^{s_1^+} {\rm d} s_1 \, \frac{|\mathcal{M}|^2}{256 \pi^3 M^3},
\end{align}
where the four-momenta are assigned as
\begin{equation}
  H^\pm(P)\to W^\pm(k)\, \bar b(p_1)\, b(p_2),
\end{equation}
such that $s_1=(p_1+p_2)^2$ and $s_2=(p_2+k)^2$. The integration limits for the phase space integration are thus given by
\begin{equation}\begin{split}
  s_2^- = &\ (m_W+m_b)^2 ,\\
  s_2^+ = &\ (M-m_b)^2,\\
  s_1^\pm = &\ 2m_b^2 + \frac{1}{2s_2} (M^2-m_b^2-s_2)(s_2-m_W^2+m_b^2)\\ &\qquad \pm \frac{1}{2s_2} \lambda^{\frac{1}{2}} (s_2, m_b^2, m_W^2) \lambda^{\frac{1}{2}} (M^2, s_2, m_b^2).
\end{split}\end{equation}
The matrix element corresponding to the $H^+\to t^*\bar b\to W^+ b\bar b$ process reads
\begin{equation}\begin{split}
  i\mathcal{M} =&\ -\bar{u}(p_2)\, \frac{g}{\sqrt{2}} V_{tb}^*\, \slashed{\epsilon}^*(k)\, P_L\, \frac{\slashed p_2+\slashed k+m_t}{(p_2+k)^2-m_t^2}\, \\
  &\qquad \times \frac{\sqrt{2}}{v} V_{tb} \big(\bar{m}_t \xi_t P_L + \bar{m}_b \xi_b P_R\big)\,  v(p_1),
\end{split}\end{equation}
where $u(p_2)$ and $v(p_1)$ denote the spinors associated with the final-state $b$ quark and $\bar b$ antiquark, respectively, $\epsilon^\mu(k)$ is the $W$-boson polarization vector, and $g$ the weak $SU(2)_L$ coupling. After spin and colour sums, the squared matrix element can be expressed as
\begin{align} 
  |\mathcal{M}|^2 =\frac{24 G_F^2 |V_{tb}|^4}{(s_2-m_t^2)^2} \left(A \xi_t^2 +2 B \xi_t \xi_b + C \xi_b^2\right),
\end{align}
where the coefficients $A$, $B$ and $C$ read
\begin{equation}\begin{split}
  A =&\ m_t^2 \bar{m}_t^2 \Big[-M^2 (m_W^2 -s_2)+m_W^2 (2s_1 +s_2)-m_b^4\\
     &\  -s_2(s_1+s_2)-m_b^2 (M^2+3m_W^2-s_1-2s_2)\Big],\\
  B =&\ m_b m_t \bar{m}_b \bar{m}_t \Big[(2m_W^4-m_W^2 s_2 -s_2^2)\\
     &\  -m_b^2(m_W^2-2s_2)-m_b^4\Big],\\
  C =&\ \bar{m}_b^2 \big[-2M^2 m_W^2 (m_W^2-s_2)+2m_W^4 s_2 -m_b^6\\
    &\ + m_b^2\big\{M^2(m_W^2\!-\!s_2) \!+\! 2m_W^4 \!+\! m_W^2 s_2 \!-\! s_2(s_1\!+\!s_2)\big\}  \\
    &\ -2m_W^2 s_2 (s_1+s_2)+s_1 s_2^2 +m_b^4(M^2-m_W^2+2s_2)\big].
\end{split}\end{equation}


\bibliographystyle{JHEP}
\bibliography{v0}

@article{ATLAS:2024kxj,
    author = "Aad, Georges and others",
    collaboration = "ATLAS",
    title = "{Climbing to the Top of the ATLAS 13 TeV data}",
    eprint = "2404.10674",
    archivePrefix = "arXiv",
    primaryClass = "hep-ex",
    reportNumber = "CERN-EP-2024-099",
    doi = "10.1016/j.physrep.2024.12.004",
    journal = "Phys. Rept.",
    volume = "1116",
    pages = "127--183",
    year = "2025"
}

@article{CMS:2024irj,
    author = "Hayrapetyan, Aram and others",
    collaboration = "CMS",
    title = "{Review of top quark mass measurements in CMS}",
    eprint = "2403.01313",
    archivePrefix = "arXiv",
    primaryClass = "hep-ex",
    reportNumber = "CMS-TOP-23-003, CERN-EP-2024-005",
    doi = "10.1016/j.physrep.2024.12.002",
    journal = "Phys. Rept.",
    volume = "1115",
    pages = "116--218",
    year = "2025"
}

@article{Rizzo:1989ci,
    author = "Rizzo, Thomas G.",
    title = "{Top Quark Decay in Models with Higgs Triplets}",
    reportNumber = "MAD/PH/531",
    doi = "10.1103/PhysRevD.41.1504",
    journal = "Phys. Rev. D",
    volume = "41",
    pages = "1504",
    year = "1990"
}

@article{CoarasaPerez:1998sqz,
    author = "Coarasa Perez, Jose Antonio and Guasch, Jaume and Sola, Joan and Hollik, Wolfgang",
    title = "{Top quark decay into charged Higgs boson in a general two Higgs doublet model: Implications for the Tevatron data}",
    eprint = "hep-ph/9808278",
    archivePrefix = "arXiv",
    reportNumber = "UAB-FT-450, KA-TP-14-1998",
    doi = "10.1016/S0370-2693(98)01246-5",
    journal = "Phys. Lett. B",
    volume = "442",
    pages = "326--334",
    year = "1998"
}

@article{Bejar:2000ub,
    author = "Bejar, Santi and Guasch, Jaume and Sola, Joan",
    title = "{Loop induced flavor changing neutral decays of the top quark in a general two Higgs doublet model}",
    eprint = "hep-ph/0011091",
    archivePrefix = "arXiv",
    reportNumber = "UAB-FT-491, KA-TP-22-2000",
    doi = "10.1016/S0550-3213(01)00044-X",
    journal = "Nucl. Phys. B",
    volume = "600",
    pages = "21--38",
    year = "2001"
}

@article{ALEPH:2013htx,
    author = "Abbiendi, G. and others",
    collaboration = "ALEPH, DELPHI, L3, OPAL, LEP",
    title = "{Search for Charged Higgs bosons: Combined Results Using LEP Data}",
    eprint = "1301.6065",
    archivePrefix = "arXiv",
    primaryClass = "hep-ex",
    reportNumber = "CERN-PH-EP-2012-369",
    doi = "10.1140/epjc/s10052-013-2463-1",
    journal = "Eur. Phys. J. C",
    volume = "73",
    pages = "2463",
    year = "2013"
}

@article{CDF:2009efz,
    author = "Aaltonen, T. and others",
    collaboration = "CDF",
    title = "{Search for charged Higgs bosons in decays of top quarks in p anti-p collisions at $\sqrt{s}$ = 1.96\,TeV}",
    eprint = "0907.1269",
    archivePrefix = "arXiv",
    primaryClass = "hep-ex",
    reportNumber = "FERMILAB-PUB-09-343-E",
    doi = "10.1103/PhysRevLett.103.101803",
    journal = "Phys. Rev. Lett.",
    volume = "103",
    pages = "101803",
    year = "2009"
}

@article{D0:2009oou,
    author = "Abazov, V. M. and others",
    collaboration = "D0",
    title = "{Search for Charged Higgs Bosons in Top Quark Decays}",
    eprint = "0908.1811",
    archivePrefix = "arXiv",
    primaryClass = "hep-ex",
    reportNumber = "FERMILAB-PUB-09-393-E",
    doi = "10.1016/j.physletb.2009.11.016",
    journal = "Phys. Lett. B",
    volume = "682",
    pages = "278--286",
    year = "2009"
}

@article{CMS:2019bfg,
    author = "Sirunyan, Albert M and others",
    collaboration = "CMS",
    title = "{Search for charged Higgs bosons in the H$^{\pm}$ $\to$ $\tau^{\pm}\nu_\tau$ decay channel in proton-proton collisions at $\sqrt{s} =$ 13 TeV}",
    eprint = "1903.04560",
    archivePrefix = "arXiv",
    primaryClass = "hep-ex",
    reportNumber = "CMS-HIG-18-014, CERN-EP-2019-025",
    doi = "10.1007/JHEP07(2019)142",
    journal = "JHEP",
    volume = "07",
    pages = "142",
    year = "2019"
}

@article{ATLAS:2024hya,
    author = "Aad, Georges and others",
    collaboration = "ATLAS",
    title = "{Search for charged Higgs bosons produced in top-quark decays or in association with top quarks and decaying via $H${\ensuremath{\pm}}{\textrightarrow}{\ensuremath{\tau}}{\ensuremath{\pm}}{\ensuremath{\nu}}{\ensuremath{\tau}} in 13\,TeV pp collisions with the ATLAS detector}",
    eprint = "2412.17584",
    archivePrefix = "arXiv",
    primaryClass = "hep-ex",
    reportNumber = "CERN-EP-2024-311",
    doi = "10.1103/PhysRevD.111.072006",
    journal = "Phys. Rev. D",
    volume = "111",
    number = "7",
    pages = "072006",
    year = "2025"
}

@article{CMS:2020osd,
    author = "Sirunyan, Albert M and others",
    collaboration = "CMS",
    title = "{Search for a light charged Higgs boson in the H$^\pm$ $\to $ cs channel in proton-proton collisions at $\sqrt{s} =$ 13 TeV}",
    eprint = "2005.08900",
    archivePrefix = "arXiv",
    primaryClass = "hep-ex",
    reportNumber = "CMS-HIG-18-021, CERN-EP-2020-057",
    doi = "10.1103/PhysRevD.102.072001",
    journal = "Phys. Rev. D",
    volume = "102",
    number = "7",
    pages = "072001",
    year = "2020"
}

@article{ATLAS:2024oqu,
    author = "Aad, Georges and others",
    collaboration = "ATLAS",
    title = "{Search for a light charged Higgs boson in $t \rightarrow H^{\pm } b$ decays, with $H^{\pm } \rightarrow cs$, in $pp$ collisions at $\sqrt{s}={13}\hbox { TeV}$ with the ATLAS detector}",
    eprint = "2407.10096",
    archivePrefix = "arXiv",
    primaryClass = "hep-ex",
    reportNumber = "CERN-EP-2024-185",
    doi = "10.1140/epjc/s10052-024-13715-4",
    journal = "Eur. Phys. J. C",
    volume = "85",
    number = "2",
    pages = "153",
    year = "2025"
}

@article{ATLAS:2023bzb,
    author = "Aad, Georges and others",
    collaboration = "ATLAS",
    title = "{Search for a light charged Higgs boson in $t \rightarrow H^{\pm}b$ decays, with $H^{\pm} \rightarrow cb$, in the lepton+jets final state in proton-proton collisions at $\sqrt{s}=13$ TeV with the ATLAS detector}",
    eprint = "2302.11739",
    archivePrefix = "arXiv",
    primaryClass = "hep-ex",
    reportNumber = "CERN-EP-2022-207",
    doi = "10.1007/JHEP09(2023)004",
    journal = "JHEP",
    volume = "09",
    pages = "004",
    year = "2023"
}

@article{ATLAS:2021upq,
    author = "Aad, Georges and others",
    collaboration = "ATLAS",
    title = "{Search for charged Higgs bosons decaying into a top quark and a bottom quark at $ \sqrt{\mathrm{s}} $ = 13 TeV with the ATLAS detector}",
    eprint = "2102.10076",
    archivePrefix = "arXiv",
    primaryClass = "hep-ex",
    reportNumber = "CERN-EP-2021-004",
    doi = "10.1007/JHEP06(2021)145",
    journal = "JHEP",
    volume = "06",
    pages = "145",
    year = "2021"
}

@article{CMS:2025plw,
    author = "Hayrapetyan, Aram and others",
    collaboration = "CMS",
    title = "{Search for charged Higgs bosons decaying into top and bottom quarks in lepton+jets final states in proton-proton collisions at $\sqrt{s}$ = 13 TeV}",
    eprint = "2512.24471",
    archivePrefix = "arXiv",
    primaryClass = "hep-ex",
    reportNumber = "CMS-B2G-24-008, CERN-EP-2025-297",
    month = "12",
    year = "2025"
}

@book{Gunion:1989we,
    author = "Gunion, John F. and Haber, Howard E. and Kane, Gordon L. and Dawson, Sally",
    title = "{The Higgs Hunter's Guide}",
    reportNumber = "SCIPP-89/13, UCD-89-4, BNL-41644",
    doi = "10.1201/9780429496448",
    isbn = "978-0-429-49644-8",
    volume = "80",
    year = "2000"
}

@article{Branco:2011iw,
    author = "Branco, G. C. and Ferreira, P. M. and Lavoura, L. and Rebelo, M. N. and Sher, Marc and Silva, Joao P.",
    title = "{Theory and phenomenology of two-Higgs-doublet models}",
    eprint = "1106.0034",
    archivePrefix = "arXiv",
    primaryClass = "hep-ph",
    doi = "10.1016/j.physrep.2012.02.002",
    journal = "Phys. Rept.",
    volume = "516",
    pages = "1--102",
    year = "2012"
}

@article{LHCHiggsCrossSectionWorkingGroup:2013rie,
    author = "Andersen, J R and others",
    editor = "Heinemeyer, S and Mariotti, C and Passarino, G and Tanaka, R",
    collaboration = "LHC Higgs Cross Section Working Group",
    title = "{Handbook of LHC Higgs Cross Sections: 3. Higgs Properties}",
    eprint = "1307.1347",
    archivePrefix = "arXiv",
    primaryClass = "hep-ph",
    reportNumber = "CERN-2013-004",
    doi = "10.5170/CERN-2013-004",
    month = "7",
    year = "2013"
}

@article{Akeroyd:2016ymd,
    author = "Akeroyd, A. G. and others",
    title = "{Prospects for charged Higgs searches at the LHC}",
    eprint = "1607.01320",
    archivePrefix = "arXiv",
    primaryClass = "hep-ph",
    reportNumber = "CERN-TH-2016-152",
    doi = "10.1140/epjc/s10052-017-4829-2",
    journal = "Eur. Phys. J. C",
    volume = "77",
    number = "5",
    pages = "276",
    year = "2017"
}

@article{Arbey:2017gmh,
    author = "Arbey, A. and Mahmoudi, F. and Stal, O. and Stefaniak, T.",
    title = "{Status of the Charged Higgs Boson in Two Higgs Doublet Models}",
    eprint = "1706.07414",
    archivePrefix = "arXiv",
    primaryClass = "hep-ph",
    reportNumber = "CERN-TH-2017-137, SCIPP-17-07",
    doi = "10.1140/epjc/s10052-018-5651-1",
    journal = "Eur. Phys. J. C",
    volume = "78",
    number = "3",
    pages = "182",
    year = "2018"
}

@article{Cheung:2022ndq,
    author = "Cheung, Kingman and Jueid, Adil and Kim, Jinheung and Lee, Soojin and Lu, Chih-Ting and Song, Jeonghyeon",
    title = "{Comprehensive study of the light charged Higgs boson in the type-I two-Higgs-doublet model}",
    eprint = "2201.06890",
    archivePrefix = "arXiv",
    primaryClass = "hep-ph",
    reportNumber = "KIAS-Q22002",
    doi = "10.1103/PhysRevD.105.095044",
    journal = "Phys. Rev. D",
    volume = "105",
    number = "9",
    pages = "095044",
    year = "2022"
}

@article{Ashanujjaman:2024lnr,
    author = "Ashanujjaman, Saiyad and Banik, Sumit and Coloretti, Guglielmo and Crivellin, Andreas and Maharathy, Siddharth P. and Mellado, Bruce",
    title = "{Anatomy of the real Higgs triplet model}",
    eprint = "2411.18618",
    archivePrefix = "arXiv",
    primaryClass = "hep-ph",
    reportNumber = "TTP24-044, P3H-24-089, PSI-PR-24-25, ZU-TH 59/24, ICPP-88",
    doi = "10.1007/JHEP04(2025)003",
    journal = "JHEP",
    volume = "04",
    pages = "003",
    year = "2025"
}

@article{Ashanujjaman:2025una,
    author = "Ashanujjaman, Saiyad and Crivellin, Andreas and Maharathy, Siddharth P. and Mellado, Bruce",
    title = "{Searching for a charged Higgs boson in top-quark decays via the $WZ$ mode}",
    eprint = "2509.07094",
    archivePrefix = "arXiv",
    primaryClass = "hep-ph",
    reportNumber = "TTP25-028, P3H-25-059, ZU-TH 53/25, ICPP-97",
    doi = "10.1103/n2g2-gld5",
    journal = "Phys. Rev. D",
    volume = "113",
    number = "11",
    pages = "115048",
    year = "2026"
}

@article{Crivellin:2021ubm,
    author = "Crivellin, Andreas and Fang, Yaquan and Fischer, Oliver and Bhattacharya, Srimoy and Kumar, Mukesh and Malwa, Elias and Mellado, Bruce and Rapheeha, Ntsoko and Ruan, Xifeng and Sha, Qiyu",
    title = "{Accumulating evidence for the associated production of a new Higgs boson at the LHC}",
    eprint = "2109.02650",
    archivePrefix = "arXiv",
    primaryClass = "hep-ph",
    reportNumber = "ICPP-057, PSI-PR-21-21, ZU-TH 38/21, CERN-TH-2021-129, LTH 1267",
    doi = "10.1103/PhysRevD.108.115031",
    journal = "Phys. Rev. D",
    volume = "108",
    number = "11",
    pages = "115031",
    year = "2023"
}

@article{Bhattacharya:2023lmu,
    author = "Bhattacharya, Srimoy and Coloretti, Guglielmo and Crivellin, Andreas and Dahbi, Salah-Eddine and Fang, Yaquan and Kumar, Mukesh and Mellado, Bruce",
    title = "{Growing Excesses of New Scalars at the Electroweak Scale}",
    eprint = "2306.17209",
    archivePrefix = "arXiv",
    primaryClass = "hep-ph",
    reportNumber = "PSI-PR-23-21, ZU-TH 31/23, ICPP-70",
    month = "6",
    year = "2023"
}

@article{Ashanujjaman:2024pky,
    author = "Ashanujjaman, Saiyad and Banik, Sumit and Coloretti, Guglielmo and Crivellin, Andreas and Maharathy, Siddharth P. and Mellado, Bruce",
    title = "{Explaining the {\ensuremath{\gamma}}{\ensuremath{\gamma}}+$X$ excesses at {\ensuremath{\approx}}151.5\,GeV via the Drell-Yan production of a Higgs triplet}",
    eprint = "2402.00101",
    archivePrefix = "arXiv",
    primaryClass = "hep-ph",
    reportNumber = "PSI-PR-24-06, ZU-TH 09/24, ICPP-79",
    doi = "10.1016/j.physletb.2025.139298",
    journal = "Phys. Lett. B",
    volume = "862",
    pages = "139298",
    year = "2025"
}

@article{Bhattacharya:2025rfr,
    author = "Bhattacharya, Srimoy and Lieberman, Benjamin and Kumar, Mukesh and Crivellin, Andreas and Fang, Yaquan and Mazini, Rachid and Mellado, Bruce",
    title = "{Emerging Excess Consistent with a Narrow Resonance at 152\,GeV in High-Energy Proton-Proton Collisions}",
    eprint = "2503.16245",
    archivePrefix = "arXiv",
    primaryClass = "hep-ph",
    reportNumber = "ICPP-94, ZU-TH 18/25",
    month = "3",
    year = "2025"
}

@article{Banik:2023vxa,
    author = "Banik, Sumit and Coloretti, Guglielmo and Crivellin, Andreas and Mellado, Bruce",
    title = "{Uncovering new Higgses in the LHC analyses of differential $ t\overline{t} $ cross sections}",
    eprint = "2308.07953",
    archivePrefix = "arXiv",
    primaryClass = "hep-ph",
    reportNumber = "PSI-PR-23-30, ZU-TH 46/23, ICPP-72",
    doi = "10.1007/JHEP01(2025)155",
    journal = "JHEP",
    volume = "01",
    pages = "155",
    year = "2025"
}

@article{Coloretti:2023yyq,
    author = "Coloretti, Guglielmo and Crivellin, Andreas and Mellado, Bruce",
    title = "{Combined explanation of LHC multilepton, diphoton, and top-quark excesses}",
    eprint = "2312.17314",
    archivePrefix = "arXiv",
    primaryClass = "hep-ph",
    reportNumber = "PSI-PR-24-01, ZU-TH 01/24, ICPP-78",
    doi = "10.1103/PhysRevD.110.073001",
    journal = "Phys. Rev. D",
    volume = "110",
    number = "7",
    pages = "073001",
    year = "2024"
}

@article{Crivellin:2024uhc,
    author = "Crivellin, Andreas and Ashanujjaman, Saiyad and Banik, Sumit and Coloretti, Guglielmo and Maharathy, Siddharth P. and Mellado, Bruce",
    title = "{Growing evidence for a Higgs triplet}",
    eprint = "2404.14492",
    archivePrefix = "arXiv",
    primaryClass = "hep-ph",
    reportNumber = "ZU-TH 24/24, PSI-PR-24-11, ICPP-81",
    doi = "10.1088/1674-1137/adbb5a",
    journal = "Chin. Phys. C",
    volume = "49",
    number = "5",
    pages = "053107",
    year = "2025"
}

@article{Czakon:2011xx,
    author = "Czakon, Michal and Mitov, Alexander",
    title = "{Top++: A Program for the Calculation of the Top-Pair Cross-Section at Hadron Colliders}",
    eprint = "1112.5675",
    archivePrefix = "arXiv",
    primaryClass = "hep-ph",
    reportNumber = "CERN-PH-TH-2011-303, TTK-11-58",
    doi = "10.1016/j.cpc.2014.06.021",
    journal = "Comput. Phys. Commun.",
    volume = "185",
    pages = "2930",
    year = "2014"
}

@article{ATLAS:2024aht,
    author = "Aad, Georges and others",
    collaboration = "ATLAS",
    title = "{Measurement of $ t\overline{t} $ production in association with additional b-jets in the e{\ensuremath{\mu}} final state in proton{\textendash}proton collisions at $ \sqrt{s} $ = 13 TeV with the ATLAS detector}",
    eprint = "2407.13473",
    archivePrefix = "arXiv",
    primaryClass = "hep-ex",
    reportNumber = "CERN-EP-2024-191",
    doi = "10.1007/JHEP01(2025)068",
    journal = "JHEP",
    volume = "01",
    pages = "068",
    year = "2025"
}

@article{Alwall:2014hca,
    author = "Alwall, J. and Frederix, R. and Frixione, S. and Hirschi, V. and Maltoni, F. and Mattelaer, O. and Shao, H. -S. and Stelzer, T. and Torrielli, P. and Zaro, M.",
    title = "{The automated computation of tree-level and next-to-leading order differential cross sections, and their matching to parton shower simulations}",
    eprint = "1405.0301",
    archivePrefix = "arXiv",
    primaryClass = "hep-ph",
    reportNumber = "CERN-PH-TH-2014-064, CP3-14-18, LPN14-066, MCNET-14-09, ZU-TH-14-14",
    doi = "10.1007/JHEP07(2014)079",
    journal = "JHEP",
    volume = "07",
    pages = "079",
    year = "2014"
}

@article{Frederix:2018nkq,
    author = "Frederix, R. and Frixione, S. and Hirschi, V. and Pagani, D. and Shao, H. -S. and Zaro, M.",
    title = "{The automation of next-to-leading order electroweak calculations}",
    eprint = "1804.10017",
    archivePrefix = "arXiv",
    primaryClass = "hep-ph",
    reportNumber = "Nikhef/2018-015, TUM-HEP-1138/18, NIKHEF-2018-015, TUM-HEP-1138-18",
    doi = "10.1007/JHEP11(2021)085",
    journal = "JHEP",
    volume = "07",
    pages = "185",
    year = "2018",
    note = "[Erratum: JHEP 11, 085 (2021)]"
}

@article{NNPDF:2017mvq,
    author = "Ball, Richard D. and others",
    collaboration = "NNPDF",
    title = "{Parton distributions from high-precision collider data}",
    eprint = "1706.00428",
    archivePrefix = "arXiv",
    primaryClass = "hep-ph",
    reportNumber = "EDINBURGH-2017-08, NIKHEF-2017-006, OUTP-17-04P, TIF-UNIMI-2017-3, CAVENDISH-HEP-17-06, CERN-TH-2017-077, Edinburgh 2017/08,
  Nikhef/2017-006, OUTP-17-04P,TIF-UNIMI-2017-3",
    doi = "10.1140/epjc/s10052-017-5199-5",
    journal = "Eur. Phys. J. C",
    volume = "77",
    number = "10",
    pages = "663",
    year = "2017"
}

@article{TheATLAScollaboration:2014rfk,
    title = "{ATLAS Pythia 8 tunes to 7 TeV data}",
    reportNumber = "ATL-PHYS-PUB-2014-021",
    month = "11",
    year = "2014"
}

@article{Cacciari:2008gp,
    author = "Cacciari, Matteo and Salam, Gavin P. and Soyez, Gregory",
    title = "{The anti-$k_t$ jet clustering algorithm}",
    eprint = "0802.1189",
    archivePrefix = "arXiv",
    primaryClass = "hep-ph",
    reportNumber = "LPTHE-07-03",
    doi = "10.1088/1126-6708/2008/04/063",
    journal = "JHEP",
    volume = "04",
    pages = "063",
    year = "2008"
}

@article{Cacciari:2011ma,
    author = "Cacciari, Matteo and Salam, Gavin P. and Soyez, Gregory",
    title = "{FastJet User Manual}",
    eprint = "1111.6097",
    archivePrefix = "arXiv",
    primaryClass = "hep-ph",
    reportNumber = "CERN-PH-TH-2011-297",
    doi = "10.1140/epjc/s10052-012-1896-2",
    journal = "Eur. Phys. J. C",
    volume = "72",
    pages = "1896",
    year = "2012"
}

@article{Bahl:2025you,
    author = "Bahl, Henning and Kumar, Romal and Weiglein, Georg",
    title = "{Impact of interference effects on Higgs-boson searches in the di-top final state at the LHC}",
    eprint = "2503.02705",
    archivePrefix = "arXiv",
    primaryClass = "hep-ph",
    reportNumber = "DESY-25-030",
    doi = "10.1007/JHEP05(2025)098",
    journal = "JHEP",
    volume = "05",
    pages = "098",
    year = "2025"
}

@article{Bevilacqua:2021cit,
    author = "Bevilacqua, Giuseppe and Bi, Huan-Yu and Hartanto, Heribertus Bayu and Kraus, Manfred and Lupattelli, Michele and Worek, Malgorzata",
    title = "{$ t\overline{t}b\overline{b} $ at the LHC: on the size of corrections and b-jet definitions}",
    eprint = "2105.08404",
    archivePrefix = "arXiv",
    primaryClass = "hep-ph",
    reportNumber = "TTK-21-16, P3H-21-029, CAVENDISH-HEP-21/08",
    doi = "10.1007/JHEP08(2021)008",
    journal = "JHEP",
    volume = "08",
    pages = "008",
    year = "2021"
}

@article{Bevilacqua:2022twl,
    author = "Bevilacqua, Giuseppe and Bi, Huan-Yu and Hartanto, Heribertus Bayu and Kraus, Manfred and Lupattelli, Michele and Worek, Malgorzata",
    title = "{tt{\textasciimacron}bb{\textasciimacron} at the LHC: On the size of off-shell effects and prompt b-jet identification}",
    eprint = "2202.11186",
    archivePrefix = "arXiv",
    primaryClass = "hep-ph",
    reportNumber = "TTK-22-06, P3H-22-012, CAVENDISH-HEP-22/02",
    doi = "10.1103/PhysRevD.107.014028",
    journal = "Phys. Rev. D",
    volume = "107",
    number = "1",
    pages = "014028",
    year = "2023"
}

@article{Hocker:2001xe,
    author = "Hocker, Andreas and Lacker, H. and Laplace, S. and Le Diberder, F.",
    title = "{A New approach to a global fit of the CKM matrix}",
    eprint = "hep-ph/0104062",
    archivePrefix = "arXiv",
    reportNumber = "LAL-01-06",
    doi = "10.1007/s100520100729",
    journal = "Eur. Phys. J. C",
    volume = "21",
    pages = "225--259",
    year = "2001"
}

@article{ParticleDataGroup:2024cfk,
    author = "Navas, S. and others",
    collaboration = "Particle Data Group",
    title = "{Review of particle physics}",
    doi = "10.1103/PhysRevD.110.030001",
    journal = "Phys. Rev. D",
    volume = "110",
    number = "3",
    pages = "030001",
    year = "2024"
}

@article{Huang:2020rtx,
    author = "Huang, Xu-Dong and Wu, Xing-Gang and Zeng, Jun and Yu, Qing and Zheng, Xu-Chang and Xu, Shuai",
    title = "{Determination of the top-quark $\overline{MS}$ running mass via its perturbative relation to the on-shell mass with the help of the principle of maximum conformality}",
    eprint = "2005.04996",
    archivePrefix = "arXiv",
    primaryClass = "hep-ph",
    doi = "10.1103/PhysRevD.101.114024",
    journal = "Phys. Rev. D",
    volume = "101",
    number = "11",
    pages = "114024",
    year = "2020"
}

@article{Aparisi:2021tym,
    author = "Aparisi, Javier and others",
    title = "{$m_b$ at $m_H$: The Running Bottom Quark Mass and the Higgs Boson}",
    eprint = "2110.10202",
    archivePrefix = "arXiv",
    primaryClass = "hep-ph",
    doi = "10.1103/PhysRevLett.128.122001",
    journal = "Phys. Rev. Lett.",
    volume = "128",
    number = "12",
    pages = "122001",
    year = "2022"
}

@article{Ma:2024xeq,
    author = "Ma, Shun-Yue and Huang, Xu-Dong and Zheng, Xu-Chang and Wu, Xing-Gang",
    title = "{Precise Determination of the Bottom-Quark On-Shell Mass Using Its Four-Loop Relation to the MS{\textasciimacron} -Scheme Running Mass}",
    eprint = "2406.18025",
    archivePrefix = "arXiv",
    primaryClass = "hep-ph",
    doi = "10.1088/0256-307X/41/10/101201",
    journal = "Chin. Phys. Lett.",
    volume = "41",
    number = "10",
    pages = "101201",
    year = "2024"
}

@article{Shen:2022yuo,
    author = "Shen, Xiao-Min and Hu, YaLu and Sun, ChuanLe and Gao, Jun",
    title = "{Decay of the charged Higgs boson and the top quark in two-Higgs-doublet model at NNLO in QCD}",
    eprint = "2201.08139",
    archivePrefix = "arXiv",
    primaryClass = "hep-ph",
    doi = "10.1007/JHEP05(2022)157",
    journal = "JHEP",
    volume = "05",
    pages = "157",
    year = "2022"
}

@article{Chen:2022wit,
    author = "Chen, Long-Bin and Li, Hai Tao and Wang, Jian and Wang, Yefan",
    title = "{Analytic result for the top-quark width at next-to-next-to-leading order in QCD}",
    eprint = "2212.06341",
    archivePrefix = "arXiv",
    primaryClass = "hep-ph",
    doi = "10.1103/PhysRevD.108.054003",
    journal = "Phys. Rev. D",
    volume = "108",
    number = "5",
    pages = "054003",
    year = "2023"
}

@article{Crivellin:2013wna,
    author = "Crivellin, Andreas and Kokulu, Ahmet and Greub, Christoph",
    title = "{Flavor-phenomenology of two-Higgs-doublet models with generic Yukawa structure}",
    eprint = "1303.5877",
    archivePrefix = "arXiv",
    primaryClass = "hep-ph",
    doi = "10.1103/PhysRevD.87.094031",
    journal = "Phys. Rev. D",
    volume = "87",
    number = "9",
    pages = "094031",
    year = "2013"
}

@article{CMS:2014mxl,
    author = "Khachatryan, Vardan and others",
    collaboration = "CMS",
    title = "{Measurement of the Ratio $\mathcal B(t \to Wb)/\mathcal B(t \to Wq)$ in pp Collisions at $\sqrt{s}$ = 8\,TeV}",
    eprint = "1404.2292",
    archivePrefix = "arXiv",
    primaryClass = "hep-ex",
    reportNumber = "CMS-TOP-12-035, CERN-PH-EP-2014-052",
    doi = "10.1016/j.physletb.2014.06.076",
    journal = "Phys. Lett. B",
    volume = "736",
    pages = "33--57",
    year = "2014"
}

@article{Misiak:2017bgg,
    author = "Misiak, Mikolaj and Steinhauser, Matthias",
    title = "{Weak radiative decays of the $B$ meson and bounds on $M_{H^\pm }$ in the Two-Higgs-Doublet Model}",
    eprint = "1702.04571",
    archivePrefix = "arXiv",
    primaryClass = "hep-ph",
    reportNumber = "TTP17-004, IFT-1-2017",
    doi = "10.1140/epjc/s10052-017-4776-y",
    journal = "Eur. Phys. J. C",
    volume = "77",
    number = "3",
    pages = "201",
    year = "2017"
}

@article{Jung:2010ik,
    author = "Jung, Martin and Pich, Antonio and Tuzon, Paula",
    title = "{Charged-Higgs phenomenology in the Aligned two-Higgs-doublet model}",
    eprint = "1006.0470",
    archivePrefix = "arXiv",
    primaryClass = "hep-ph",
    doi = "10.1007/JHEP11(2010)003",
    journal = "JHEP",
    volume = "11",
    pages = "003",
    year = "2010"
}

@article{Enomoto:2015wbn,
    author = "Enomoto, Tetsuya and Watanabe, Ryoutaro",
    title = "{Flavor constraints on the Two Higgs Doublet Models of Z$_{2}$ symmetric and aligned types}",
    eprint = "1511.05066",
    archivePrefix = "arXiv",
    primaryClass = "hep-ph",
    reportNumber = "OU-HET-882, CTPU-15-17",
    doi = "10.1007/JHEP05(2016)002",
    journal = "JHEP",
    volume = "05",
    pages = "002",
    year = "2016"
}

@article{Coutinho:2024zyp,
    author = "Coutinho, Antonio M. and Karan, Anirban and Miralles, V{\'\i}ctor and Pich, Antonio",
    title = "{Light scalars within the $ \mathcal{CP} $-conserving Aligned-two-Higgs-doublet model}",
    eprint = "2412.14906",
    archivePrefix = "arXiv",
    primaryClass = "hep-ph",
    doi = "10.1007/JHEP02(2025)057",
    journal = "JHEP",
    volume = "02",
    pages = "057",
    year = "2025"
}

@article{Fuks:2024qdt,
    author = "Fuks, Benjamin and Goodsell, Mark D. and Murphy, Taylor",
    title = "{Monojets from compressed weak frustrated dark matter}",
    eprint = "2409.03014",
    archivePrefix = "arXiv",
    primaryClass = "hep-ph",
    doi = "10.1103/PhysRevD.111.055010",
    journal = "Phys. Rev. D",
    volume = "111",
    number = "5",
    pages = "055010",
    year = "2025"
}

@article{Buckley:2014ana,
    author = {Buckley, Andy and Ferrando, James and Lloyd, Stephen and Nordstr{\"o}m, Karl and Page, Ben and R{\"u}fenacht, Martin and Sch{\"o}nherr, Marek and Watt, Graeme},
    title = "{LHAPDF6: parton density access in the LHC precision era}",
    eprint = "1412.7420",
    archivePrefix = "arXiv",
    primaryClass = "hep-ph",
    reportNumber = "GLAS-PPE-2014-05, MCNET-14-29, IPPP-14-111, DCPT-14-222",
    doi = "10.1140/epjc/s10052-015-3318-8",
    journal = "Eur. Phys. J. C",
    volume = "75",
    pages = "132",
    year = "2015"
}

@article{Bierlich:2022pfr,
    author = "Bierlich, Christian and others",
    title = "{A comprehensive guide to the physics and usage of PYTHIA 8.3}",
    eprint = "2203.11601",
    archivePrefix = "arXiv",
    primaryClass = "hep-ph",
    reportNumber = "LU-TP 22-16, MCNET-22-04, FERMILAB-PUB-22-227-SCD",
    doi = "10.21468/SciPostPhysCodeb.8",
    journal = "SciPost Phys. Codeb.",
    volume = "2022",
    pages = "8",
    year = "2022"
}

@article{Alwall:2014bza,
    author = {Alwall, Johan and Duhr, Claude and Fuks, Benjamin and Mattelaer, Olivier and \"Ozt\"urk, Deniz Gizem and Shen, Chia-Hsien},
    title = "{Computing decay rates for new physics theories with FeynRules  and MadGraph 5\_aMC@NLO}",
    eprint = "1402.1178",
    archivePrefix = "arXiv",
    primaryClass = "hep-ph",
    reportNumber = "CERN-PH-TH-2014-020, MCNET-14-03, IPPP-14-10, DCPT-14-20, CP3-14-11, CALT-68-2877, ZU-TH-02-14",
    doi = "10.1016/j.cpc.2015.08.031",
    journal = "Comput. Phys. Commun.",
    volume = "197",
    pages = "312--323",
    year = "2015"
}

@article{Artoisenet:2012st,
    author = "Artoisenet, Pierre and Frederix, Rikkert and Mattelaer, Olivier and Rietkerk, Robbert",
    title = "{Automatic spin-entangled decays of heavy resonances in Monte Carlo simulations}",
    eprint = "1212.3460",
    archivePrefix = "arXiv",
    primaryClass = "hep-ph",
    reportNumber = "NIKHEF-2012-021, CERN-PH-TH-2012-329",
    doi = "10.1007/JHEP03(2013)015",
    journal = "JHEP",
    volume = "03",
    pages = "015",
    year = "2013"
}

@article{Darme:2023jdn,
    author = "Darm\'e, Luc and others",
    title = "{UFO 2.0: the \textquoteleft{}Universal Feynman Output\textquoteright{} format}",
    eprint = "2304.09883",
    archivePrefix = "arXiv",
    primaryClass = "hep-ph",
    reportNumber = "BONN-TH-2023-03, DESY-23-051, FERMILAB-PUB-23-138-T, KA-TP-06-2023,
  MCNET-23-06, P3H-23-023, TIF-UNIMI-2023-11",
    doi = "10.1140/epjc/s10052-023-11780-9",
    journal = "Eur. Phys. J. C",
    volume = "83",
    number = "7",
    pages = "631",
    year = "2023"
}

@article{Christensen:2009jx,
      author         = "Christensen, Neil D. and others",
      title          = "{A Comprehensive approach to new physics simulations}",
      journal        = "Eur. Phys. J.",
      volume         = "C71",
      year           = "2011",
      pages          = "1541",
      doi            = "10.1140/epjc/s10052-011-1541-5",
      eprint         = "0906.2474",
      archivePrefix  = "arXiv",
      primaryClass   = "hep-ph",
      reportNumber   = "CP3-09-24, HD-THEP-09-11, IPHC-PHENO-09-01,
                        MSUHEP-090612, NIKHEF-2009-009",
      SLACcitation   = "%%CITATION = ARXIV:0906.2474;%%"
}

@article{Alloul:2013bka,
    author = "Alloul, Adam and Christensen, Neil D. and Degrande, C{\'e}line and Duhr, Claude and Fuks, Benjamin",
    title = "{FeynRules  2.0 - A complete toolbox for tree-level phenomenology}",
    eprint = "1310.1921",
    archivePrefix = "arXiv",
    primaryClass = "hep-ph",
    reportNumber = "CERN-PH-TH-2013-239, MCNET-13-14, IPPP-13-71, DCPT-13-142, PITT-PACC-1308",
    doi = "10.1016/j.cpc.2014.04.012",
    journal = "Comput. Phys. Commun.",
    volume = "185",
    pages = "2250--2300",
    year = "2014"
}

@article{Degrande:2011ua,
    author = "Degrande, Celine and Duhr, Claude and Fuks, Benjamin and Grellscheid, David and Mattelaer, Olivier and Reiter, Thomas",
    title = "{UFO - The Universal FeynRules Output}",
    eprint = "1108.2040",
    archivePrefix = "arXiv",
    primaryClass = "hep-ph",
    reportNumber = "CP3-11-25, IPHC-PHENO-11-04, IPPP-11-39, DCPT-11-78, MPP-2011-68",
    doi = "10.1016/j.cpc.2012.01.022",
    journal = "Comput. Phys. Commun.",
    volume = "183",
    pages = "1201--1214",
    year = "2012"
}

@article{Jezo:2018yaf,
    author = "Je{\v{z}}o, Tom{\'a}{\v{s}} and Lindert, Jonas M. and Moretti, Niccolo and Pozzorini, Stefano",
    title = "{New NLOPS predictions for $t \bar{t} +b$ -jet production at the LHC}",
    eprint = "1802.00426",
    archivePrefix = "arXiv",
    primaryClass = "hep-ph",
    reportNumber = "IPPP-18-7, ZU-TH-6-18",
    doi = "10.1140/epjc/s10052-018-5956-0",
    journal = "Eur. Phys. J. C",
    volume = "78",
    number = "6",
    pages = "502",
    year = "2018"
}

@article{Dicus:1987fk,
    author = "Dicus, Duane A. and Willenbrock, Scott S. D.",
    title = "{Photon Pair Production and the Intermediate Mass Higgs Boson}",
    reportNumber = "MAD/PH/389",
    doi = "10.1103/PhysRevD.37.1801",
    journal = "Phys. Rev. D",
    volume = "37",
    pages = "1801",
    year = "1988"
}

@article{Dicus:1994bm,
    author = "Dicus, D. and Stange, A. and Willenbrock, S.",
    title = "{Higgs decay to top quarks at hadron colliders}",
    eprint = "hep-ph/9404359",
    archivePrefix = "arXiv",
    reportNumber = "CPP-94-18, BNL-60339, ILL-TH-94-9",
    doi = "10.1016/0370-2693(94)91017-0",
    journal = "Phys. Lett. B",
    volume = "333",
    pages = "126--131",
    year = "1994"
}

@article{Flacke:2025dwk,
    author = "Flacke, Thomas and Fuks, Benjamin and Kim, Dongchan and Kim, Jinheung and Lee, Seung J. and Munoz-Aillaud, L{\'e}andre",
    title = "{New physics in toponium's shadow?}",
    eprint = "2512.03220",
    archivePrefix = "arXiv",
    primaryClass = "hep-ph",
    reportNumber = "KIAS-Q25020",
    doi = "10.1016/j.physletb.2026.140613",
    journal = "Phys. Lett. B",
    volume = "879",
    pages = "140613",
    year = "2026"
}

@article{Bernreuther:1997gs,
    author = "Bernreuther, W. and Flesch, M. and Haberl, P.",
    title = "{Signatures of Higgs bosons in the top quark decay channel at hadron colliders}",
    eprint = "hep-ph/9709284",
    archivePrefix = "arXiv",
    reportNumber = "PITHA-97-34",
    doi = "10.1103/PhysRevD.58.114031",
    journal = "Phys. Rev. D",
    volume = "58",
    pages = "114031",
    year = "1998"
}

@article{Gaemers:1984sj,
    author = "Gaemers, K. J. F. and Hoogeveen, F.",
    title = "{Higgs Production and Decay Into Heavy Flavors With the Gluon Fusion Mechanism}",
    reportNumber = "ITFA-84-6",
    doi = "10.1016/0370-2693(84)91711-8",
    journal = "Phys. Lett. B",
    volume = "146",
    pages = "347--349",
    year = "1984"
}
\end{document}